\documentclass[journal,10p]{IEEEtran}

\usepackage{cite}
\usepackage[hyphens]{url}
\usepackage[utf8]{inputenc}
\usepackage{graphicx}
\usepackage{float}
\usepackage{color, colortbl}
\usepackage{xcolor}
\usepackage{array}
\usepackage{multirow}
\usepackage{footnote}
\usepackage[normal]{threeparttable}
\usepackage{mathtools}

\usepackage{adjustbox}
\usepackage{subfig}
\usepackage{siunitx}

\usepackage{booktabs} 
\usepackage{comment} 
\usepackage{rotating}
\usepackage{tablefootnote}
\usepackage{amssymb}
\usepackage{makecell}
\usepackage{breakcites}
\usepackage[utf8]{inputenc} 
\usepackage[T1]{fontenc} 

\usepackage{amsmath,amsfonts,amsthm} 
\usepackage{tikz} 
\usepackage{tikz-cd}
\usepackage{gensymb}
\usepackage{siunitx}
\usepackage[]{multirow}
\usepackage{array,multirow}
\usepackage{xfrac}
\usetikzlibrary{positioning,arrows} 
\usepackage{tikzsymbols} 
\usepackage{blindtext} 

\usepackage{array}

\usepackage{graphicx}
\usepackage{subfig}

\makeatletter
\newcommand{\removelatexerror}{\let\@latex@error\@gobble}
\makeatother

\makeatletter

\newcommand{\Rmnum}[1]{\expandafter\@slowromancap\romannumeral #1@}
\makeatother
\usepackage{graphicx,epsfig}
\usepackage{amsmath,amssymb,amsfonts,nccmath}
\usepackage{mathrsfs}
\usepackage{algorithm}
\usepackage{algorithmicx}
\usepackage{algpseudocode}
\usepackage{ulem}
\usepackage{diagbox}
\usepackage{authblk}
\usepackage{comment}
\usepackage{cleveref}

\newcommand*{\depaddr}[1]{\dagmark Computer and Information Sciences, \asmark Electrical and Computer Engineering} 

\newcommand*{\instaddr}[1]{\dagmark University of Delaware~~~~~ \ddagmark Cisco Systems}

\newcommand*{\dagmark}[1][\dag]{\textsuperscript{\dag}}
\newcommand*{\asmark}[1][*]{\textsuperscript{*}}
\newcommand*{\ddagmark}[1][\ddag]{\textsuperscript{\ddag}}
\newcommand*{\email}[1]{\texttt{#1}}

\begin{document}

\title{Traffic Characteristics of Extended Reality}

\author{%
Abdullah Alnajim\dagmark[1]~~~ Seyedmohammad Salehi\dagmark[1]~~~ Chien-Chung Shen\dagmark[1]~~~ Malcolm Smith\ddagmark[1]\\
\instaddr{\dagmark[1]}  \\
\email{\{alnajim,salehi,cshen\}@udel.edu, \{mmsmith\}@cisco.com}
}

\maketitle


\begin{abstract}
This tutorial paper analyzes the traffic characteristics of immersive experiences with extended reality (XR) technologies, including Augmented reality (AR), virtual reality (VR), and mixed reality (MR). The current trend in XR applications is to offload the computation and rendering to an external server and use wireless communications between the XR head-mounted display (HMD) and the access points. This paradigm becomes essential owing to (1) its high flexibility (in terms of user mobility) compared to remote rendering through a wired connection, and (2) the high computing power available on the server compared to local rendering (on HMD). The requirements to facilitate a pleasant XR experience are analyzed in three aspects: capacity (throughput), latency, and reliability. For capacity, two VR experiences are analyzed: a human eye-like experience and an experience with the Oculus Quest 2 HMD. For latency, the key components of the motion-to-photon (MTP) delay are discussed. For reliability, the maximum packet loss rate (or the minimum packet delivery rate) is studied for different XR scenarios. Specifically, the paper reviews optimization techniques that were proposed to reduce the latency, conserve the bandwidth, extend the scalability, and/or increase the reliability to satisfy the stringent requirements of the emerging XR applications.
\end{abstract}
\section{Introduction} 
Augmented reality (AR), virtual reality (VR), and mixed reality (MR), together termed extended reality (XR), are technologies of broad societal impacts. Through immersive experiences, these technologies bring a new paradigm for how we can interact among each other and with the world, offering unprecedented experiences and unlimited possibilities that will enhance our work and lives in many ways. For instance, XR raises productivity for enterprise users and vertical markets such as health care, manufacturing, retail, and transportation, where practitioners can visually interact with one another and/or with digital information to perform remote collaboration, diagnosis, and maintenance. For consumers, XR provides personalized content and make everyday experiences more realistic, engaging, and satisfying.

Initially, XR is mainly consumed locally and statically, where the head mounted displays (HMDs), like HTC Vive and Oculus Rift, are tethered to high-end computers equipped with GPU. Nowadays, many localized and individual XR capabilities have been moved to the edge and/or the cloud\footnote{Such as Hololens and Oculus Quest 2} to facilitate XR applications where multiple geographically separated people can communicate and interact as if they were face to face in the same location. Therefore, XR applications delivered over mobile wireless networks have attracted huge interest \cite{Bastug2017,Abari2017EnablingReality,Qualcomm2016MakingMobile}. In particular, XR has become the first wave of killer apps of the 5G ecosystem \cite{ABIResearchQualcomm2017}. 

\begin{figure*}[ht] \centering
\hspace*{-0.05in}
  \begin{tabular}{cc}
  \includegraphics[width=.48\linewidth]{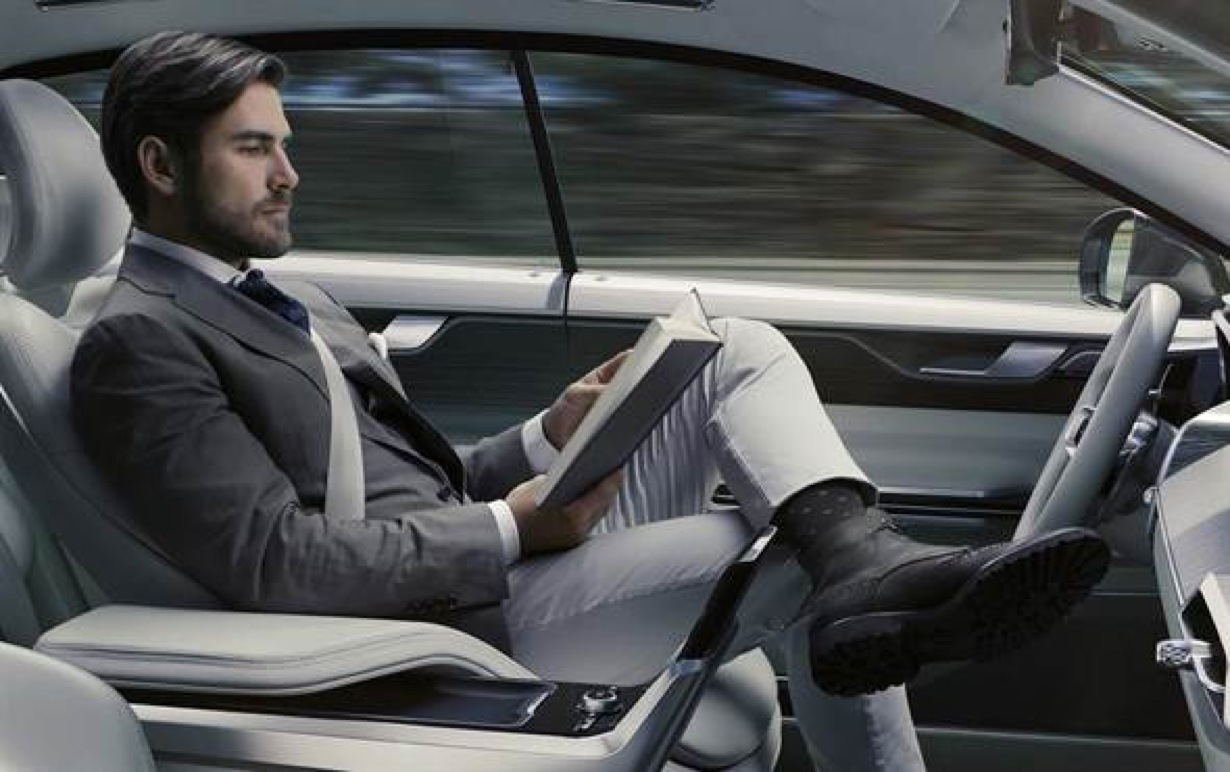} &
  \includegraphics[width=.48\linewidth]{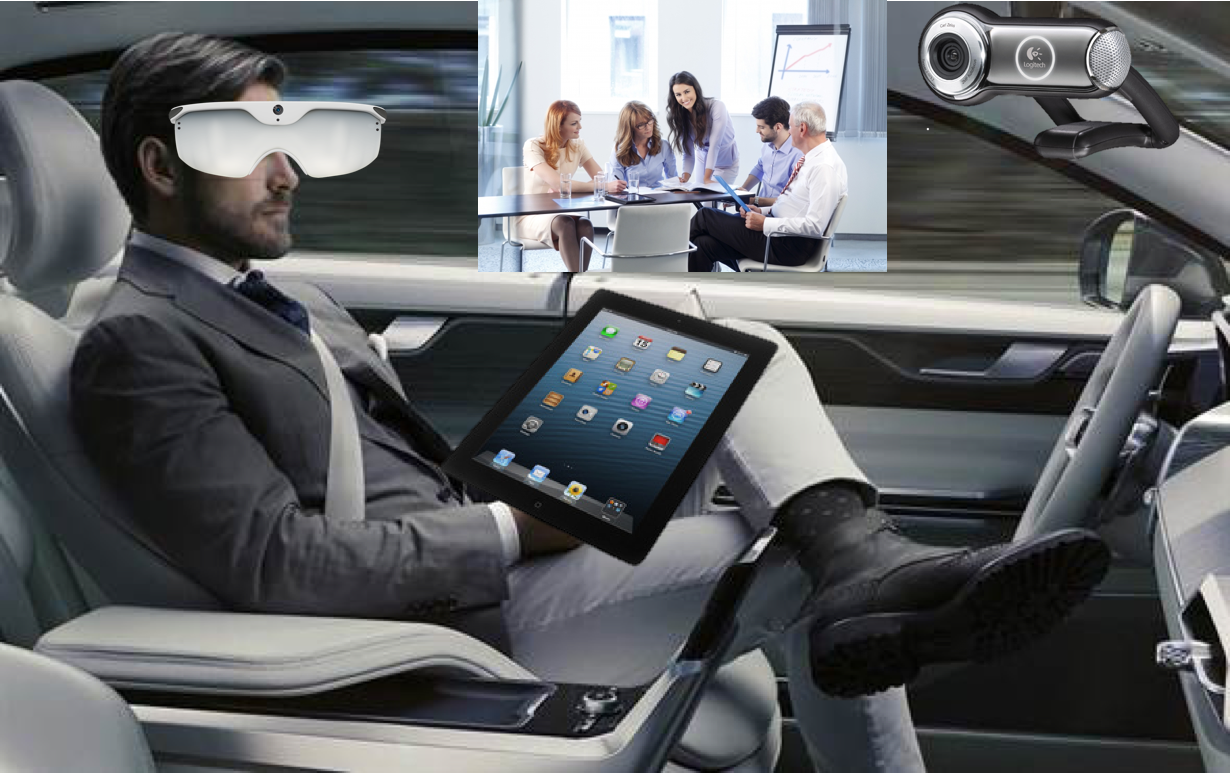} \\
  (a) Free time in self-driving car & (b) MR video conferencing \\
  \includegraphics[width=.48\linewidth]{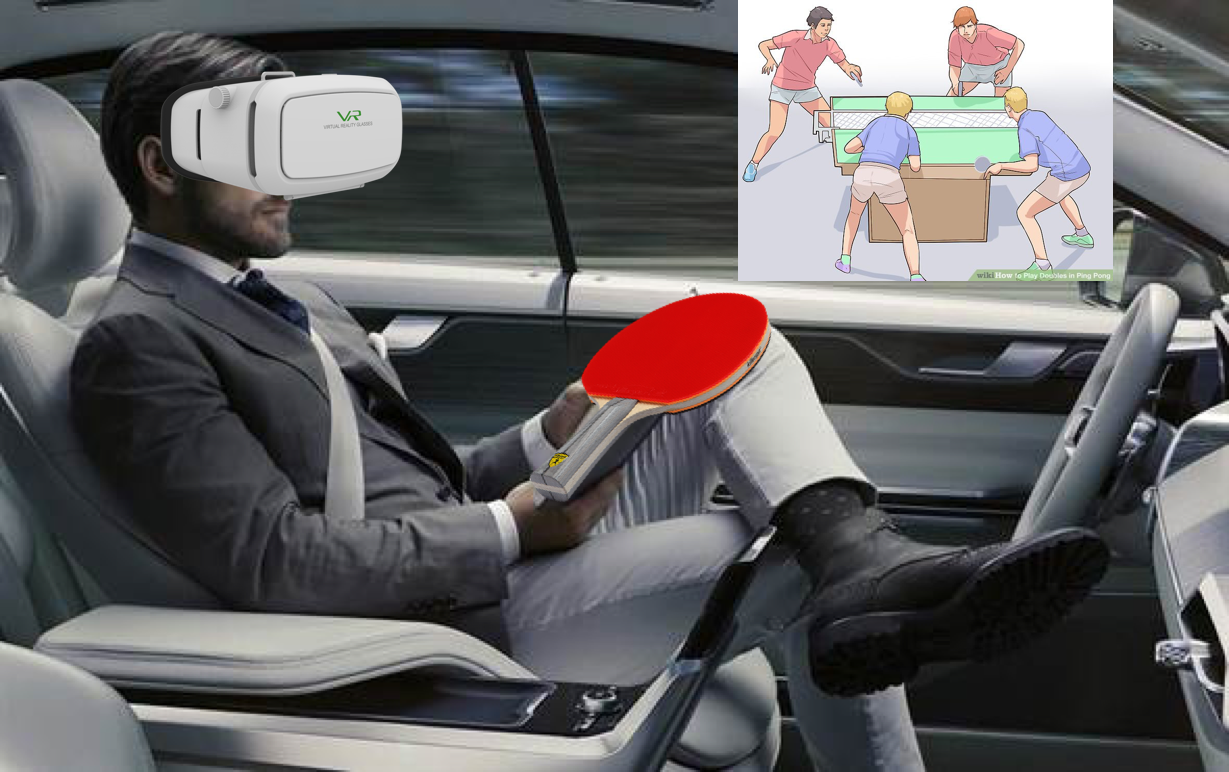} &
  \includegraphics[width=.48\linewidth]{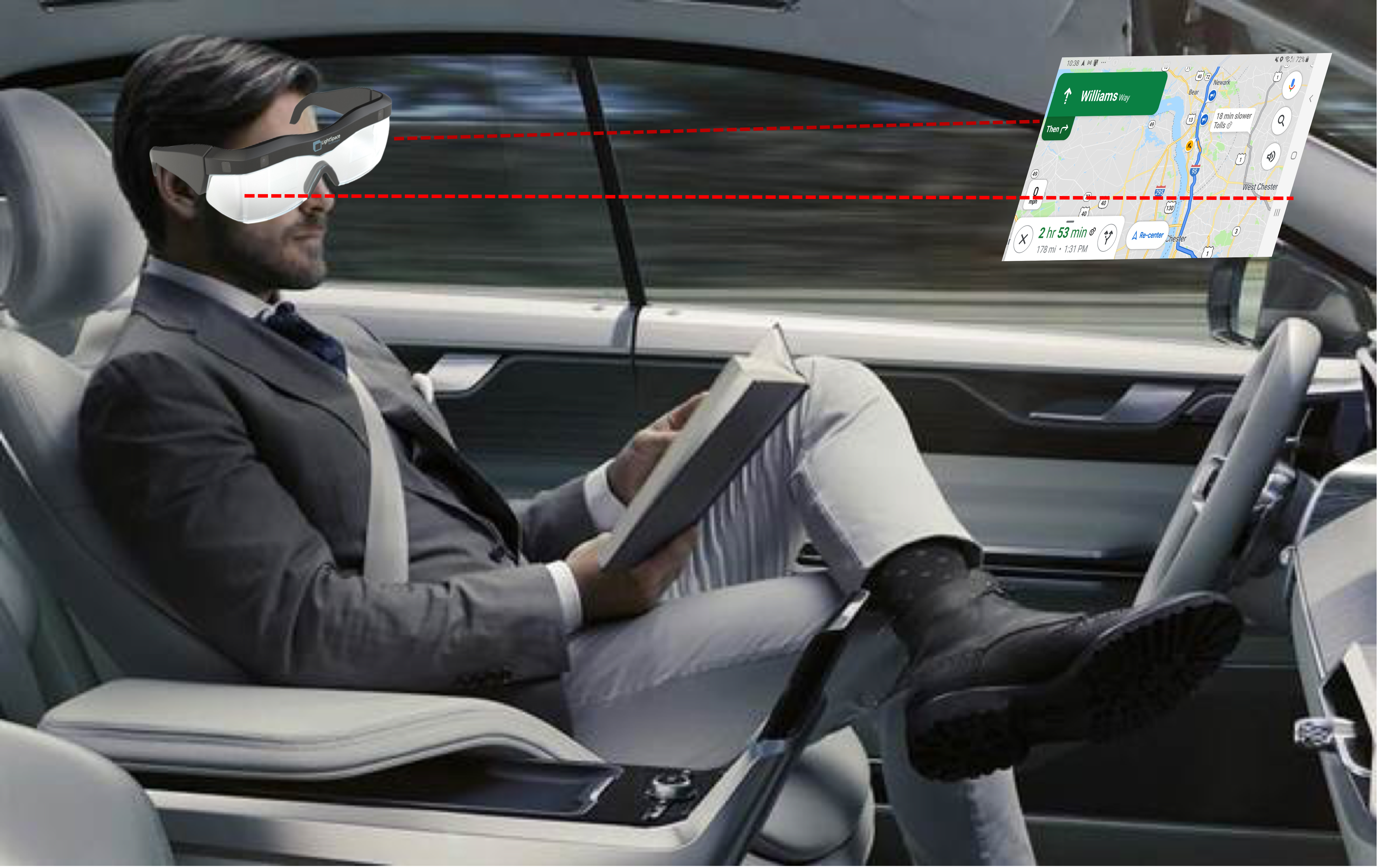}\\
  (c) VR social interactive game & (d) AR smart navigation
  \end{tabular}
  \caption{\label{fig:car}Applications of AR and VR in the context of self-driving cars (courtesy of Volvo)}
\end{figure*}

Fig. \ref{fig:car} depicts certain XR scenarios in the context of self-driving cars. Other than reading a book in an autonomous car, as depicted in Fig. \ref{fig:car}(a), by wearing a pair of MR glasses (such as Microsoft's HoloLens) and employing a video camera, you may participate in video conferencing and communicate face-to-face with remote colleagues in real-time as if we were all in the same room, 
as depicted in Fig. \ref{fig:car}(b). In contrast,
by wearing a VR goggle, you may play virtual interactive table tennis
with friends in real-time who are located in other places or self-driving cars, as shown in Fig. \ref{fig:car}(c). Fig. \ref{fig:car}(d) depicts the use of AR to check the route information and change it virtually while being able to see the physical world around you or even reading a book. 

Having light-weight wireless HMDs with immersive XR experiences poses multiple challenging requirements which can be summarized in high capacity (throughput), ultra-low latency, and ultra-high reliability and uniform experience. Achieving any two of these requirements may result in sacrificing the third; therefore, complying with all of these requirements is not a trivial task. This tutorial paper discusses these requirements in detail and analyzes the typical values for each requirement. Moreover, it reviews the papers that discussed one or more of these requirements as depicted in Table \ref{Table:requirements}. Moreover, by considering two VR experiences: an ideal eye-like VR and a practical VR with Oculus Quest 2, this paper gives approximate typical bit rate, latency, and packet loss values for each experience based on the discussed analysis. In addition, since numerous approaches have been proposed to satisfy one or more of these requirements, the paper thoroughly reviews and classifies these approaches based on multiple criteria.

\begin{table}[ht]
\centering
\begin{tabular}{|l||*{5}{c|}}\hline
\backslashbox{papers}{requirements}
&\makebox[4em]{Capacity}&\makebox[4em]{Latency}&\makebox[4em]{Reliability} \\\hline\hline
\hfil\cite{Boos2016, Kojic2019, Liu2019, Zhang2018} &&\checkmark&\\\hline
\hfil\cite{Cuervo2018, Huawei2016, Mangiante2017, Tan2018} &\checkmark&\checkmark&\\\hline
\hfil\cite{Nasrallah2019} &&\checkmark&\checkmark\\\hline
\hfil\cite{ABIResearchQualcomm2017, Adame2020, Bastug2017,  Elbamby2018, Hu2020, HuaweiiLab2017, HuaweiTechnologies2016} &\checkmark&\checkmark&\checkmark\\\hline
\end{tabular}
\captionof{table}{Summary of the works that discussed XR traffic requirements in this paper}
\label{Table:requirements}
\end{table}

Before we analyze the requirements of immersive XR experiences, we first review the two types of videos used in XR, namely: 360-degree videos and volumetric videos, and highlight the difference between them in Section \ref{Sec:VolumetricVideos}. Then, in Section \ref{Sec:Capacity} the capacity requirement of XR is scrutinized by showcasing two VR experiences: an ideal eye-like VR and a practical VR with Oculus Quest 2. After that, the main components of the motion-to-photon latency are elaborated, and the typical latencies for both weak-interaction and strong-interaction VR services are investigated in Section \ref{Sec:Latency}. Section \ref{Sec:Optimizations} reviews and classifies the efforts that have been done to minimize the required bandwidth and/or the MTP latency. In Section \ref{Sec:Reliability}, the paper analyzes the typical maximum allowable packet loss rate. Finally, Section \ref{Sec:Conclusion} concludes the paper and highlights directions of future work.
\section{360-degree Videos vs. Volumetric Videos}
\label{Sec:VolumetricVideos}

\begin{figure*}[htpb!] 
    \centering 
\includegraphics[width=0.8\linewidth]{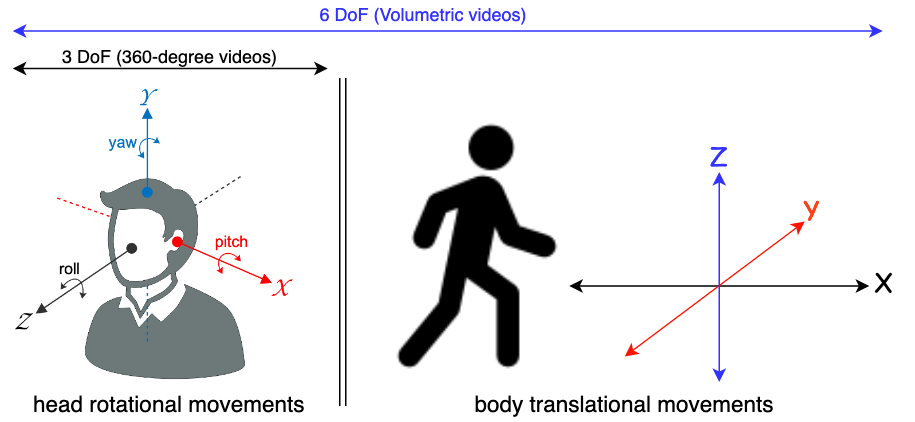} 
    \caption{There are two types of movements, head rotational movements (left), and body translational movements (right). 360-degree videos consists of the head rotational movements only, while volumetric videos consists of both the head rotational and the body translational movements} 
    \label{fig:videosTypes}
\end{figure*}

A 360-degree video is a spherical video where the viewer sits in the center and has three degrees of freedom (3DoF): yaw with 360 degrees, pitch with 180 degrees, and roll.  In contrast, a volumetric video allows the viewer to have 6DoF, in addition to enabling 3D rendering \cite{Qian2019}. These two features of volumetric videos (6DoF and 3D rendering) make them highly immersive, interactive, and expressive. In volumetric videos, the viewers are allowed to change the orientation (yaw, pitch, and roll) and position (x, y, and z) of their viewports. Fig. \ref{fig:videosTypes} shows the difference between these two types of videos. Another difference is that a 360-degree video is always a sphere and can be projected to become a plane, while a volumetric video is a real 3D video and uses approaches such as 3D meshes (polygons) and Point Cloud (PtCl) to represent the video. 

In \cite{Qian2019}, the authors used PtCl to represent their volumetric video. Unlike regular video frames, where each point in a frame is represented as a pixel with three values (R[ed], G[reen], and B[lue]) to show the color, a point in a frame using PtCl is represented as a voxel with six values: three for the color (R, G, and B) and three for the position ($x$, $y$, and $z$). Each voxel needs 9 bytes: one byte for each value of the color (R, G, and B) and two bytes for each value of the position ($x$, $y$, and $z$). Based on the 6DoF AR video captured at AT\&T SHAPE\footnote{https://www.youtube.com/watch?v=C3YUwhZ7M2g} \cite{Qian2019}, an uncompressed frame is represented using 50,360 points on average. Based on that, the amount of bandwidth needed to run 30 frame-per-second (fps) video equals $50,360 \times 9 \times 8 \times 30 \approx 103.74$ Mbps, and to run 60 fps video requires twice amount of bandwidth. In comparison, a VR example of volumetric video\footnote{https://www.youtube.com/watch?v=feGGKasvamg} is also referenced in \cite{Qian2019}.  

\begin{figure*}[htpb!] 
    \centering 
\includegraphics[width= 0.9\linewidth]{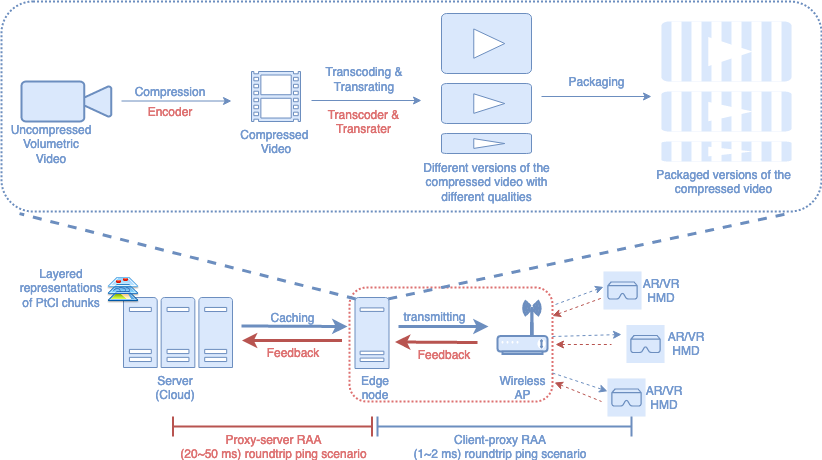} 
    \caption{Edge-based AR/VR architecture with RAA}
    \label{fig:ARVR} 
\end{figure*}

One of the main contributions of \cite{Qian2019} is the use of rate adaptation algorithms (RAAs) to dynamically adjust the quality of the video based on the quality of the channel and the transmission rate. Specifically, one RAA is deployed between the cloud server and the edge node for the voxel stream, and another RAA between the edge node and the end mobile device for the pixel stream, as depicted in Fig. \ref{fig:ARVR}. The RAA deployed between the end device and the edge node is similar to the regular RAAs used in video streaming (such as YouTube) with one key modification to accommodate the multi-viewports design which is composed of one main viewport and multiple sub-viewports likely to be seen by the viewer.
One naive approach to such an RAA design is to make all the sub-viewports of equal quality level. Another advanced method is to assign the quality of the sub-viewports based on their likelihood of being seen by the viewer. The second RAA, which is deployed between the server and the edge node, is based on a layered representation of the volumetric video in their solution. This layered representation allowed the cloud server to transmit the volumetric video based on the bandwidth between the server and the edge node. Mainly, multiple layers of each PtCl chunk are created by deleting different parts of voxels for each layer, which resulted in multiple layers ($L_0$ to $L_n$) of the video. Out of these layers, only the base layer $L_0$, which contains the minimum possible number of points, is self-contained and viewable without additional points from other layers. Other layers ($L_1$ to $L_n$) are used to increase the quality of the volumetric video by integrating the points of each layer to the points of the layers below it. Therefore, to get the quality $i$ of the volumetric video, the points of $L_0$ to $L_i$ need to be combined. The value of $i$ could be determined based on the available bandwidth between the cloud server and the edge node.

In the next section, the capacity (or bandwidth) required to transmit VR/AR traffic is reviewed.

\section{The Capacity Requirement of XR}
\label{Sec:Capacity}

Immersive XR incurs a high bandwidth consumption in the downlink (DL), the uplink (UL), or both. For instance, in a typical VR experience, the VR HMD sends small packets containing data related to the position (in case of volumetric videos) and the orientation (in case of volumetric and 360-degree videos) of the user to the edge server to retrieve the appropriate video segment with size equals to the HMD's field of view ($FoV$). Such data is typically small, which incurs low bandwidth consumption in the UL, generally in the range of several kbps \cite{Hu2020, Tan2018}. Consequently, based on the received location and orientation information, the edge server locates, renders, and transmits the target video segment back to the HMD. Such a high-resolution video segment consumes much high bandwidth in the DL. 

In contrast, in an immersive AR experience, the communications between the AR HMD and the edge server consume higher bandwidth in the UL and low bandwidth in the DL. The reason behind this is that the HMD captures the physical scene that the user is currently seeing, and sends it to the edge server. The latter identifies the targeted virtual object(s), that need to be displayed, and their locations based on the uploaded scene and sends them back to the HMD to be overlaid on top of the real scene. The amount of information of these virtual objects is typically much smaller in comparison to that of the scene.

For an immersive MR experience, where we have both complicated virtual object(s) and real scene, the bandwidth consumption for both UL and DL is high.

To achieve an eye-like VR experience, in \cite{Bastug2017}, the authors analyze the required throughput by comparing it to the real eye experience. In general, humans can process an equivalent of 5.2 gigabits of sensory input (sound and light) per second. Typically, a single eye's fovea can detect approximately 200 dots per single degree of view. Converting these dots into pixels per degree ($ppd$) depends on the size of the pixel and the distance from the display as we sill see later. A person with 20/20 vision can see up to 60 $ppd$. More details about these numbers will be discussed later in the paper.

\subsection{\bf Display quality: from resolution to pixel per degree ($ppd$)}

There are many ways to measure the performance of displays (such as XR HMD, LED monitor, and OLED TV) and the quality of digital images. Intuitively, performance or quality can be measured by resolution which is the number of pixels representing the display or the digital image. The resolution is usually shown as the number of the horizontal pixels times the number of the vertical pixels (eg., 1920 $\times$ 1080). This method is adequate to compare two displays of the same size and from the same distance. However, when the size of the two digital displays or images is different, this method becomes impractical. In other words, two digital displays with the same resolutions, but with different sizes will have different qualities. The image quality of the bigger display is less than the image quality of the smaller one, since the same number of pixels is used to stretch and cover a bigger area of the larger screen. Therefore, another way to measure the quality is the pixels density, i.e., pixels per inch ($ppi$), is introduced, where the quality is measured by the number of pixels in an inch. In this case, two different sized screens will have the same perceived quality if they have the same $ppi$. To calculate $ppi$, we use the following equation\footnote{usually the size of a display is defined by its diagonal length in inches. Therefore, the first part of the equation could be used to calculate $ppi$. Alternatively, the rest of the equation could be used to compute $ppi$ by knowing either the vertical or horizontal length of the display. The same thing is apply for $ppd$ as seen in Eq. \ref{eq:ppd_h} and \ref{eq:ppd_hV2}}:
\begin{align}
\label{eq:ppi}
ppi = \frac{D_{(pixels)}}{D_{(inches)}} =  \frac{W_{(pixels)}}{W_{(inches)}} =  \frac{H_{(pixels)}}{H_{(inches)}},
\end{align}
where $D_{(pixels)}$ is the number of pixels on the diagonal (could be computed using the Pythagorean Theorem), $D_{(inches)}$ is the diagonal length of the digital display in inches (the size of the display),  $W_{(pixels)}$ is the number of horizontal pixels, $W_{(inches)}$ is the horizontal length of the digital display in inches, $H_{(pixels)}$ is  the number of vertical pixels, and $H_{(inches)}$ is the vertical length of the digital display in inches.

\begin{figure}[htbp!]
\centering
\includegraphics[width=\linewidth]{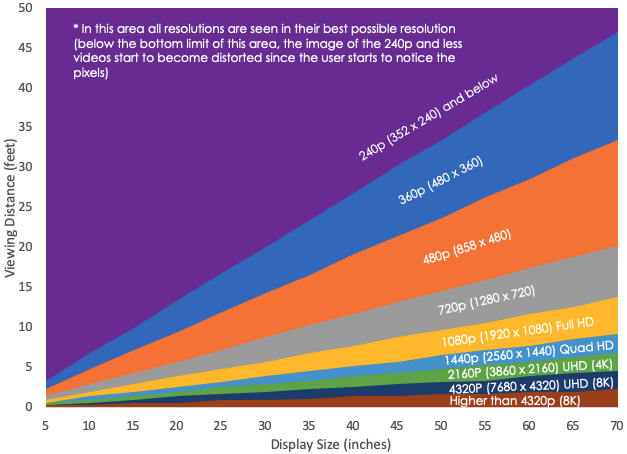} 
\caption{Display size vs. viewing distance; in the purple area, all the resolutions have the same quality}
\label{fig:displaySizeVSviewingDistance}
\end{figure}

Therefore, if we have a 40-inch display with a resolution of (1,920 pixels $\times$ 1,080 pixels), the resulting $ppi$ equals
$\frac{\sqrt{1920^2+1080^2}}{40}\approx 55 ppi$. Although this measurement is adequate to compare the image quality of two displays with different sizes, it does not consider the distance between the user and the display. The quality of an image is affected by the distance between the image and the viewer. Being closer to the display may make the viewer notice the {\em individual} pixels, which negatively affect the perceived quality of the image. In other words, two displays with the same $ppi$ but viewed from different distances will be experienced with different perceived qualities.

For any digital display,
Fig. \ref{fig:displaySizeVSviewingDistance} \cite{Bastug2017} depicts the relationship among the viewing distance, display size, and display resolution. Specifically, it tells the {\em optimal} display resolution choice subject to the given viewing distance and display size for the {\em best achievable} viewer perceived quality. The {\em optimality} can be defined in terms of either the price paid or the bandwidth consumed subject to the given viewing distance and display size. For instance, given a display size of 55", higher display resolution (i.e., higher price or bandwidth) becomes {\em necessary} when the display is viewed from a shorter distance to obtain the best achievable perceived quality. Similarly, given the same display size of 55", lower display resolution (i.e., lower price or bandwidth) becomes {\em good enough} when the display is viewed from a longer distance to obtain the best achievable perceived quality.

As a practical example, when a viewer would like to watch a 55" TV 18' away, buying a 480p resolution TV would be the optimal choice for the best achievable perceived quality. Paying more for higher resolution will not achieve any better perceived quality; getting a lower resolution TV leads to lower perceived quality. 
From a different perspective, given a viewing distance of 5', higher display resolution (up to a certain upper bound) would be necessary to accommodate bigger displace size to enjoy the best achievable perceived quality.
More generally, the purple area in the figure covers all the combinations of display size and viewing distance where resolution 240p or lower is the optimal choice for the best achievable perceived quality. 

\begin{figure*}[htpb!] 
    \centering 
\includegraphics[width=\linewidth]{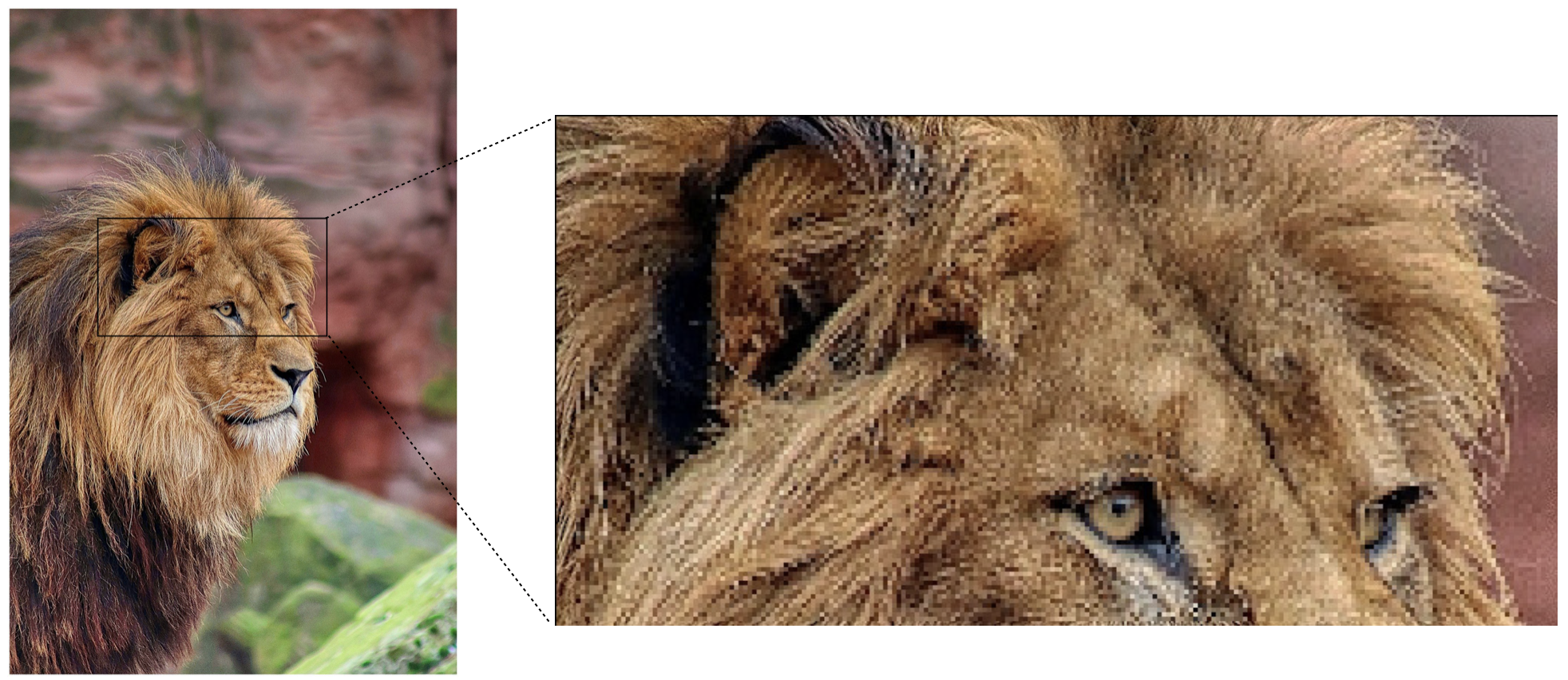} 
    \caption{The effect of distance on the perceived quality of an image: the left hand side is the original image at the original distance, and the right hand side is the viewport when the viewer moves closer} 
    \label{fig:distAffect} 
\end{figure*}

For the XR cases, although the used contents have high resolution (usually 8K or more), having a very distance of only 1 to 3 inches between the viewer's eyes and the screens of the HMDs will negatively affect the perceived quality due to the fact that the viewer can only see {\em a portion of the contents} termed {\em viewport}. Fig. \ref{fig:distAffect} demonstrates the effect of the distance on the perceived quality of a lion image. Therefore, to eliminate the effect of the distance on the perceived quality of the image (measured by $ppi$), a new measurement, namely pixels per degree ($ppd$), is introduced, which represents the number of pixels per a single degree from the eye's fovea to the image. When the distance between the image and the eyes' fovea decreased, the size of the viewable area, `corresponding to' a single degree, becomes smaller. The following elaborates the details.

$ppd$ is measured either horizontally or vertically.
The following formula calculates $ppd$:

\begin{equation}
\label{eq:ppd_h}
ppd = \frac{W_{(pixels)}}{FoV_{(h)}} = \frac{H_{(pixels)}}{FoV_{(v)}},
\end{equation}
where $W_{(pixels)}$ is the number of horizontal pixels per eye, $FoV_{(h)}$ is  the horizontal Filed of View in degrees, $H_{(pixels)}$ is the number of vertical pixels per eye, and $FoV_{(v)}$ is the vertical Filed of View in degrees.

\begin{figure}[htpb!] 
    \centering 
\includegraphics[width=\linewidth]{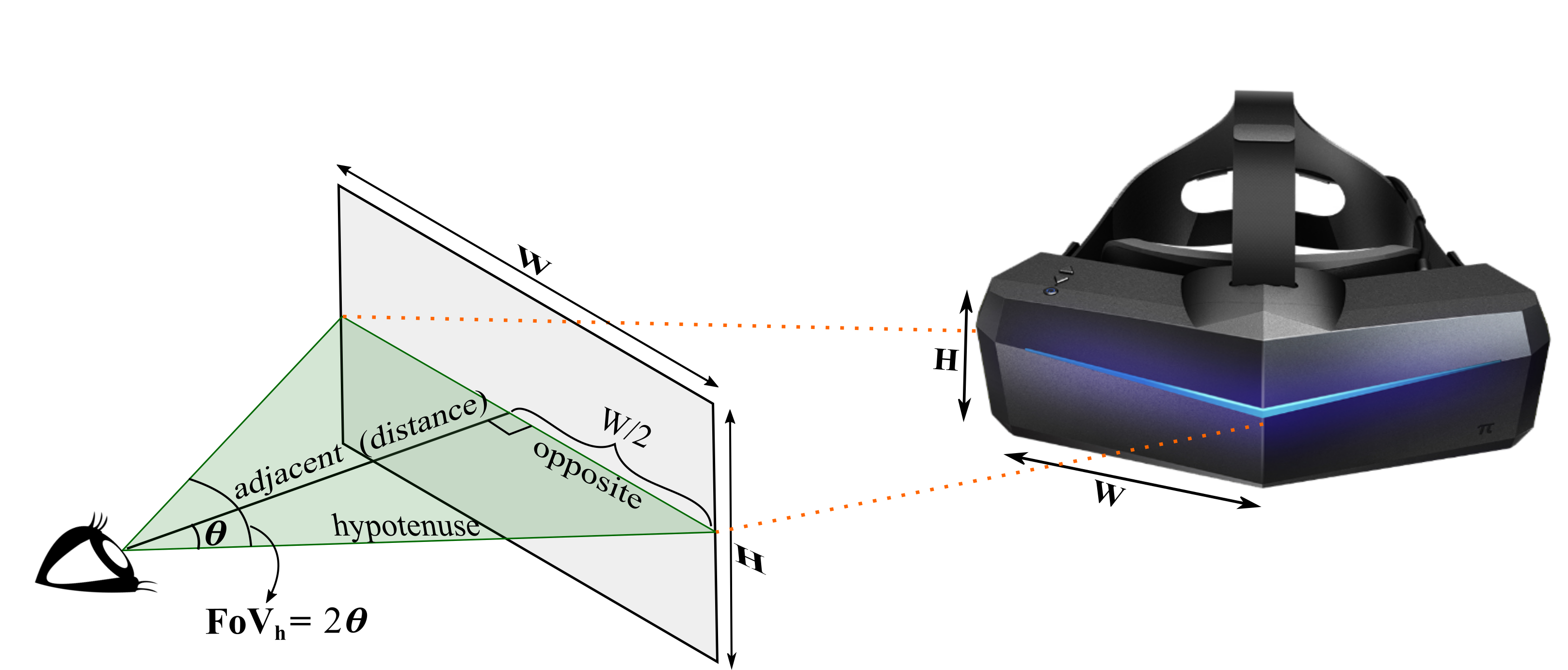} 
    \caption{Calculating the horizontal Field of View ($FoV_{(h)}$) using the display width and viewing distance} 
    \label{fig:FoV} 
\end{figure}

The field of view ($FoV$) is the maximum viewable extent a viewer can see (horizontally and vertically), which could be limited by the capability of the used device (Fig. \ref{fig:FoV}) or the human eyes (Fig. \ref{fig:HumanEyesFoV}). In Fig. \ref{fig:FoV}, the horizontal field of view ($FoV_{(h)}$) equals to 2$\theta$. 
Since the resulting triangle for computing $\theta$ is a right triangle, we have, from trigonometric ratios, that $tan(\theta)=\frac{opposite}{adjacent}$ which implies $\theta=tan^{-1}(\frac{opposite}{adjacent})$, where `opposite' represents half of the screen width in inches and `adjacent' represents the distance between the viewer's eye and the screen in inches. Thus, horizontal ($FoV_{(h)}$) and vertical ($FoV_{(v)}$) field of views can be calculated as follow: 
\begin{align}
\label{eq:FoV_h}
FoV_{(h)} = 2 \times \underbrace{tan^{-1}(\frac{0.5 \times \overbrace{W_{(inches)}}^{\frac{W_{(pixels)}}{ppi}}}{Dis_{(inches)}})}_{\theta}
\end{align}
\begin{align}
\label{eq:FoV_v}
FoV_{(v)} = 2 \times \underbrace{tan^{-1}(\frac{0.5 \times \overbrace{H_{(inches)}}^{\frac{H_{(pixels)}}{ppi}}}{Dis_{(inches)}})}_{\theta},
\end{align}
where $FoV_{(h)}$ is the horizontal Field of View in degree\footnote{if $\theta$ is in radian, it has to be converted to degree by multiply it by \ang{57.3} for both $FoV_{(h)}$ and $FoV_{(v)}$}, $W_{(inches)}$ is the width of screen in inches, $Dis_{(inches)}$ is the distance between the user's eyes and the screen in inches, $FoV_{(v)}$ is the vertical Field of View in degree, and $H_{(inches)}$ is the height of screen in inches

From Eqs. \ref{eq:ppd_h}, \ref{eq:FoV_h}, and \ref{eq:FoV_v}, we have:
\begin{align}
\label{eq:ppd_hV2}
ppd = \frac{W_{(pixels)}}{2 \times tan^{-1}(\frac{0.5 \times W_{(inches)}}{Dis_{(inches)}})} \\\notag= \frac{H_{(pixels)}}{2 \times tan^{-1}(\frac{0.5 \times H_{(inches)}}{Dis_{(inches)}})},
\end{align}
where $W_{(pixels)}$ is the number of horizontal pixels per eye, $W_{(inches)}$ is the width of screen in inches, $H_{(pixels)}$ is the number of vertical pixels per eye, $H_{(inches)}$ is the height of screen in inches, and $Dis_{(inches)}$ is the distance between the user's eyes and the screen in inches

To clarify these equations, we consider an HMD with resolution of 2,880$\times$1,600 (or 1,440$\times$1,600 per eye), a screen dimension of  5.01"$\times$5.57" (horizontal$\times$vertical), and a viewing distance of 2.5" from a viewer's eyes to the display. From Eqs. \ref{eq:ppi}, \ref{eq:FoV_h}, and \ref{eq:ppd_h},  $ppi$ equals 287 pixels per inch ($\frac{1440}{5.01}$), $FoV_{(h)}$ is approximately equal to \ang{90} ($2 \times tan^{-1}(\frac{0.5 \times 5.01}{2.5})$), and $ppd$ is equal to 16 pixels per degree ($\frac{1,440}{90}$). 
Similarly, we can use Eqs. \ref{eq:ppi}, \ref{eq:FoV_v}, and \ref{eq:ppd_h} to get 287, \ang{96}, and 16 for $ppi$, $FoV_{(v)}$, and $ppd$, respectively.

\subsection{\bf Required capacity of two VR experiences}

To derive the required capacity of VR, we consider two experiences: ideal eye-like VR and practical VR with Oculus Quest 2.

\subsubsection{\bf Ideal eye-like VR}
\label{Sec:eye-like}
The required capacity in bits per second for an ideal eye-like VR experience can be generalized from \cite{Bastug2017} as follows. 
\begin{align}
\label{eq:throghput3}
required~capacity = [2\times(FoV_{(h)}\times FoV_{(v)}\times 2~ppd)]\\\notag\times b_{(depth)} \times frame Rate \times Compression Factor^{-1},
\end{align}
where $FoV_{(h)}$ is the horizontal field of view in degrees of a single eye, $FoV_{(v)}$ is the vertical field of view in degrees of a single eye, $ppd$\footnote{As discussed in this paper, the $ppd$ value is affected by the size of and the distance from the display.} is the pixel density in pixels per degree, $b_{(depth)}$\footnote{$b_{(depth)} \in$ \{$1,2,3,4,5,6,\dots$\}, where $b_{(depth)} = 1$ represents $2^1 = 2$ colors, $b_{(depth)} = 2$ represents $2^2 = 4$ colors, {\em etc}. In today's standards, each RGB color channel is represented by 8 $\sim$ 16 bits per color (bpc) which corresponds to 8 bpc $\times$ 3 $\sim$ 16 bpc $\times$ 3 bits per pixel (bpp) when no chroma subsampling is used (a.k.a. raw file or 4:4:4 sampling). The computations in Sections \ref{Sec:eye-like}, \ref{Sec:Quest2}, and \ref{Sec:CapacityVolumetric} are under the assumption that no chroma subsampling is used.} represent the number of bits used to represent the color of a single pixel on an image, $frame Rate$\footnote{At least 60 fps for an advanced XR.} is the number of frames (or images) displayed per second (fps), $CompressionFactor$ specifies how much a compression standard reduces the bit rate. For instance, if a compression standard has a compression ratio of $x$:1, its compression factor is $x$.
To comprehend Eq. \ref{eq:throghput3}, we first review the structure of human eyes and the process of human vision, and, based on that, estimate the values of its parameters. 

A human eye consists of multiple layers. Among them is the inner layer that contains the retina which is responsible for detecting light. In turn, the retina consists of two sub-layers: the neural sub-layer that contains the photoreceptors (rods and cones), and the pigment sub-layer. The part of the retina, where we have only the pigment sub-layer, is a non-visual part of the retina, while the area, where both sub-layers exist, is the optic part of the retina. The photoreceptors are neuroepithelial cells that exist in the retina and whose main function is to convert electromagnetic waves (light energy) into electrical neural signals and send them to the brain through the optic nerve. The length of the retina is approximately 32 mm, and its peripheral area is dominated by rods that respond to the overall intensity and are responsible for vision in low light conditions (low acuity vision) \cite{Sherman2018}. Within the retina, there is a central part called the macula covered an area with a length of 5mm. The macula contains the highest density of rods and almost all the cones compared to the other regions of the retina \cite{Sherman2018}. 

Based on their responses to light, there are three types of cones corresponding to the three colors, red (R), green (G), and blue (B). It is worth to mention that each type of cones is not responsible for only a single color, but it spans over many colors; however, its wave peaks at that specific color. The authors of \cite{Sherman2018} prefer to classify them based on the peak wavelength into short (S), medium (M), and long (L) cones. Regardless of these classifications, these cones work better in highly lit places, and they are responsible for the high-acuity color vision. Moreover, the macula has a region with a length of 1.5 mm called fovea, which is the high-acuity color-vision portion of the retina since it has the highest density of cones. 
 
One of the main parameters of Eq. \ref{eq:throghput3} for computing the required capacity is $ppd$, and to compute the human eye's $ppd$, we consider only the visual acuity of the human eye, which will lead to approximate results. For every single square millimeter ($mm^2$) of the fovea, there are, on average, 147,000 cone cells \cite{Shroff2011}. Moreover, the peak density of cones at the fovea varies between 199,000 \cite{Curcio1990} and 200,000 \cite{Shroff2011} $cones/mm^2$ for an average peak, and 100,000 $cones/mm^2$ and 324,000 $cones/mm^2$ for a low and high peak \cite{Curcio1990}, respectively. Therefore, for the high peak case, we have $\sqrt{324,000}=569 ~ cones/mm$, so each cone covers $\frac{1}{569}~mm$. Since the distance between the lens and the fovea is equal to 17.1 $mm$ \cite{Serpenguzel2011}, based on Eqs. \ref{eq:FoV_h} and \ref{eq:FoV_v}, the angular view of a single cone cell (point) equals ($2 \times tan^{-1}(\frac{0.5 \times \frac{1}{569}}{17.1})\approx 0.005~degrees$), which equals ($0.005 \times 60 \approx 0.3$) arch minutes for each cell (point). If we consider each point as a pixel, there will be ($\frac{60}{0.3}=200$) points (pixels) per each degree ($ppd$) for the high peak density of cones at the fovea. Similarly, $ppd$ will be equal to 94, 133, and 120 for the low peak, average peak, and the average density of cones at the fovea, respectively. It is worth mentioning that a person with 20/20 (or 6/6 in metric system) vision acuity is able to resolve 60 $ppd$ \cite{Shroff2011}. However, the average normal young individuals have a visual acuity {\em higher} than 6/6 \cite{Blliott1995,Ohlsson2005}, usually 6/4 or better, which is equivalent to the ability to resolve 90 $ppd$ and higher. Therefore, in this paper, we consider the $ppd$ derived from the high peak retinal cones density, which is equivalent to \underline{200 $ppd$} since it represents an upper limit.  

\begin{figure}[!tbp]
  \centering
  \subfloat[Human eyes' horizontal  $FoV_{(h)}$]{\includegraphics[width=\linewidth]{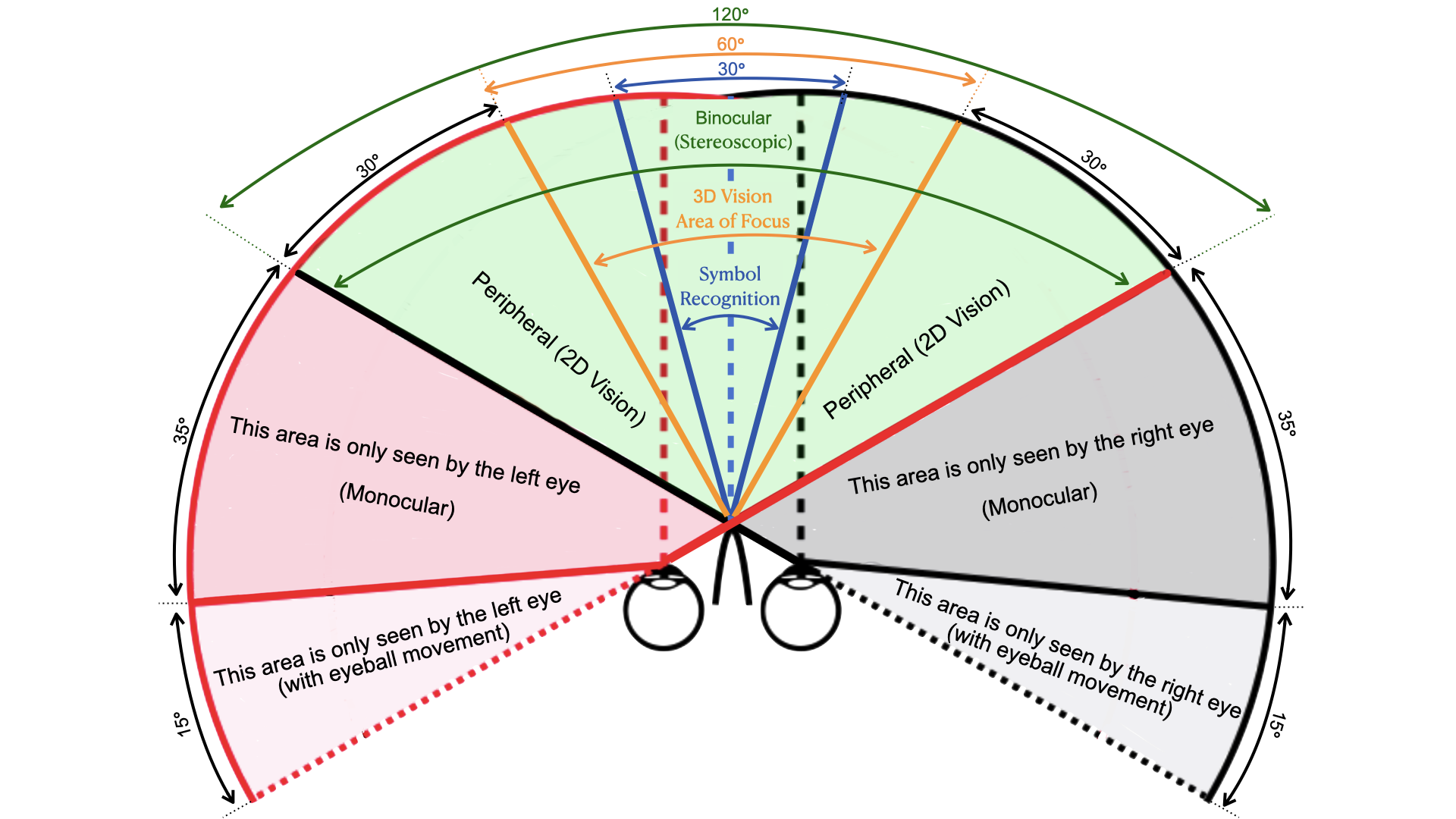}\label{fig:horizontalFoV}}
  \hfill
  \subfloat[Human eyes' vertical $FoV_{(v)}$]{\includegraphics[width=\linewidth]{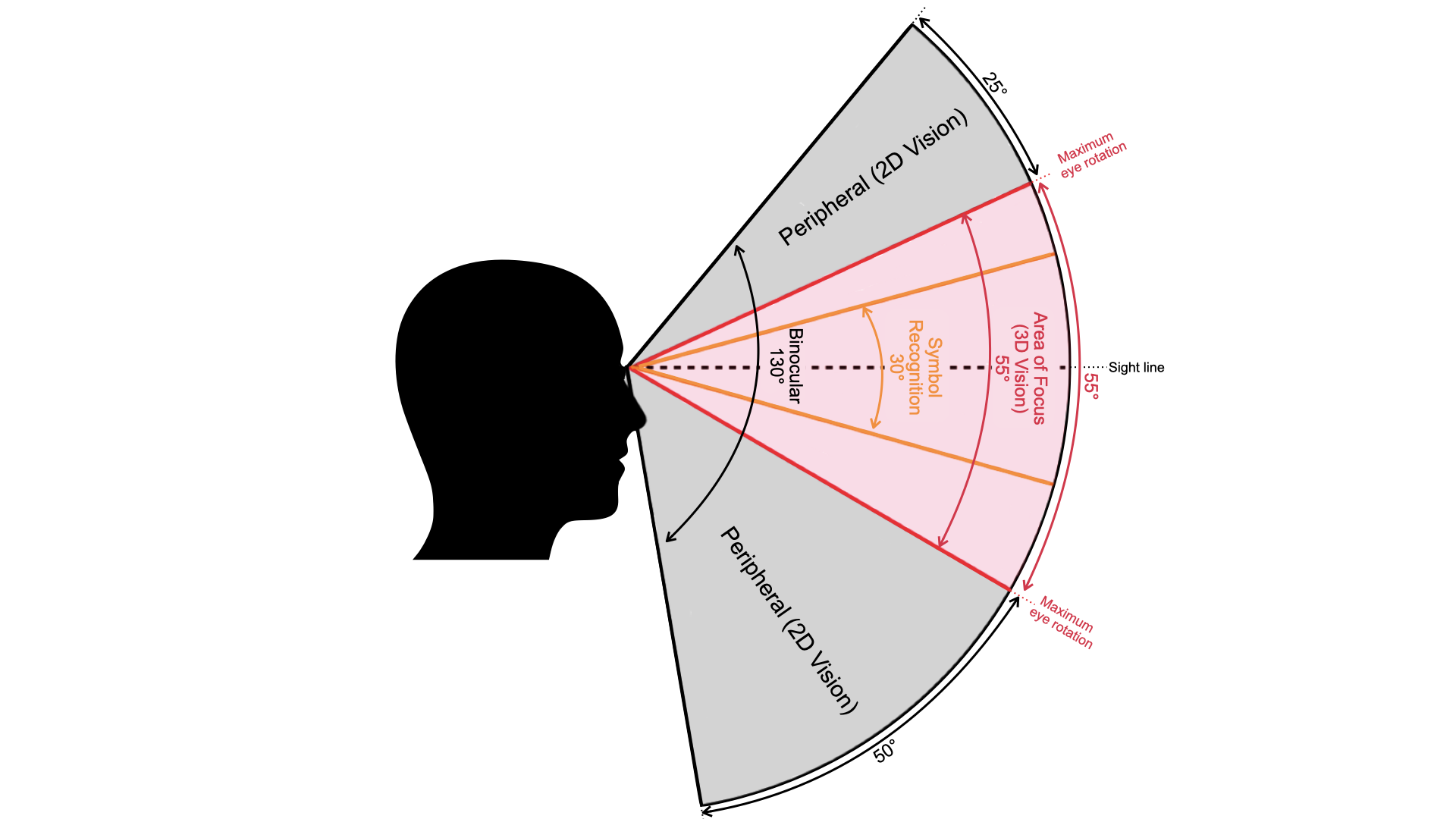}\label{fig:verticalFoV}}
  \caption{Human eyes' horizontal and vertical field of views}
  \label{fig:HumanEyesFoV}
\end{figure}

Another parameter in Eq. \ref{eq:throghput3} is $FoV$, which comprises horizontal $FoV_{(h)}$ and vertical $FoV_{(v)}$. For human eyes, $FoV$ is different from one person to another based on multiple factors, and many authors give different numbers \cite{Koslicki2017, Mazuryk1999, Sherman2018}. However, on average, $FoV_{(h)}$ for each eye is 155\degree\ \cite{Sherman2018} as shown in Fig. \ref{fig:HumanEyesFoV}(a). When a person looks straightforward, the angular view boarded by the solid red lines, started from the left eye, represents $FoV_{(h)}$ for the left eye, which is equivalent to 155\degree, and the one bounded by the solid black lines represents $FoV_{(v)}$ of the right eye. If we consider the eyeballs' movements, an additional 15\degree\ will be added for each eye, as shown by the angular view bounded by the dotted red and black lines in the edge of Fig. \ref{fig:HumanEyesFoV}(a). The green angular view of 120\degree\ in the figure represents the binocular (stereoscopic) region, which is the region seen by both eyes from two different angles of view. This region consists of multiple parts: the symbol recognition region (30\degree) bounded by the solid blue lines, the comfort zone (area of focus 60\degree) which is the 3D visual region between the solid orange lines, and the peripheral (2D Vision) region with a 30\degree\ angular view in both sides of the binocular region. 

In order to handle the Binocular region, there are two different ways used by HMDs: {\em monoscopic} and {\em stereoscopic}. In the monoscopic HMDs, the same image is fed to both lenses, while in the stereoscopic ones, a different image is fed for each lens. Since the latter is more immersive, we use the stereoscopic HMD in this case. In addition, we assume that no optimization is used to process the sent videos (i.e., the scene of the Binocular region is sent twice, one for each eye). Therefore, the resulting $FoV_{(h)}$ is \underline{155\degree} for each eye. On the other hand, Fig. \ref{fig:HumanEyesFoV}(b) shows that the vertical field of view $FoV_{(v)}$ of human eyes is \underline{130\degree} \cite{Sherman2018}, where the area of focus (3D Vision) has a 55\degree\ angular view, and the up and down peripheral (2D Vision) regions have 25\degree\ and 50\degree\ angular views, respectively.  

Based on the above calculations and the assumption of no optimizations (such as viewport prediction), to achieve an eye-like experience, every single degree of the seen picture should have 200 pixels ($200~ppd$). In addition, if we assume no head or body movements, a human eye can see a region of 155\degree\ horizontally and 130\degree\ vertically. The horizontal \ang{155} region is represented by 155 $\times$ 200 = 31000 pixels, and the vertical \ang{130} region is represented by 130 $\times$ 200 = 26000 pixels. Therefore, the full region is represented by 31000 $\times$ 26000 = 806 million pixels. Considering the head rotation (\ang{360} horizontally or \ang{180} vertically), which is the same as 360-degree videos, the full picture is represented by (360 $\times$ 200) $\times$ (180 $\times$ 200) $\approx$ 2.6 billion pixels. All of the above computation is for a single picture (frame) without translational (or body) movements (i.e., only head rotations). The difference between body translational movements and head rotational movements is depicted in Fig. \ref{fig:videosTypes}.

Since videos in general are sequences of frames (or pictures) that appear sequentially over a period of time, we need to know what is the minimum number of frames per second needed  to achieve an immersive XR experiment. 
Although 30 fps are adequate for regular videos, high-speed videos (for capturing fast-moving objects) require more than this number. Human eyes can capture videos up to 150 fps, and to avoid blurriness in such videos, at least 60-120 fps \cite{Bastug2017} are needed. Moreover, to ensure a smooth and continuous pixel-by-pixel movement for an HMD with 60 $ppd$ and a head movement speed of 30 degrees per seconds, we need at least a refresh rate of 1800 Hz \cite{Cuervo2018}. However, based on \cite{Potter2014}, human can perceive the meaning of a specific picture out of a rapid serial visual presentation (RSVP) of pictures and identify some of its details in as low as 13 $ms$, which corresponds to approximately \underline{77} fps. 

Another parameter in Eq. \ref{eq:throghput3} is the bit depth ($b_{(depth)}$) which is used to represent the colors in an image. The practical value of this parameter could be derived from the ability of human eyes. The amount of colors human eyes can distinguish is different from one person to another based on multiple factors (including the age). However, in general, human eyes can distinguish between 10 million colors \cite{Judd1975}. Therefore, the appropriate bit depth is 24 bits/pixel, which gives us approximately 17 million color variations. As mentioned before, each eye can observe up to 806 million pixels, and if we consider the 77 fps case with 24 bits/pixel, the size of the data that is generated every second by the human eye-like VR experience equals to (806 million pixels $\times$ 2 eyes $\times$ 77 fps $\times$ 24 bits) $\approx$ 346.8 Gigabytes $\approx$ 2.71 Terabits. 

In addition, although video compression incurs an additional delay to the XR experience as we will see in Section \ref{Sec:Latency}, these encoding algorithms help reducing the required bit rate dramatically. Even with the Wi-Fi standards of 802.11ax High Efficiency (HE) or Wi-Fi 6 \cite{802.11WG-WirelessLANWorkingGroup2017} (with a peak bit rate of a couple of Gbps) and 802.11be Extremely High Throughput (EHT) or Wi-Fi 7 \cite{802.11WG-WirelessLANWorkingGroup} (with a target throughput of at least 30 Gbps), using an efficient compression standard is necessary. The compression factor, which determines how much bit rate can be reduced with the used compression standard, depends on multiple factors, including video resolution and frame details. However, in general, we can use the maximum compression ratio that an encoding standard could achieve to estimate the required throughput by an immersive XR experience. For H.265 High-Efficiency Video Coding (HEVC), the maximum compression ratio is 600:1 \cite{Elbamby2018}, which means a reduction in the required bit rate by a compression factor of 600. Therefore, the required bit rate of this XR experience will be ($\frac{2.71 ~Terabits}{600}\approx$) 4.62 Gbps \cite{Bastug2017}. In 2020, a new compression standard, namely H.266 Versatile Video Coding (VVC) \cite{TheJointVideoExpertsTeamJVET2020}, is finalized, which targets XR contents, a.k.a. 360\degree ~ and volumetric videos. This new standard is anticipated to improve HEVC by up to 50\% in terms of bit rate reduction \cite{Fautier2017}. Therefore, theoretically, the required bit rate will be further reduced to become 2.31 Gbps. The performance of VVC is thoroughly investigated in \cite{Sidaty2019} and showed an improvement in the range between 31\% to 40\% over HEVC in different scenarios. 

The above computed bit rates are for different scenarios of transmitting only the viewport (i.e., the visible region of the video by the human eyes' FoV). Although transmitting only the viewport reduces the required bit rate, it needs an accurate prediction of the user's head movement to transmit the correct viewport. Alternatively, the full 360-degree video could be transmitted and only the region of viewport will be displayed on the HMD. To compute the required bit rate to transmit the full video without compression, we use Eq. \ref{eq:throghput3} with $FoV_{(h)}$ = 360\degree and $FoV_{(v)}$ = 180\degree. Therefore, the required bit rate (without compression) is equal to [(360 $\times$ 200) $\times$ (180\degree $\times$ 200\degree)] pixels $\times$ 77 fps $\times$ 24 bits 
$\approx$ 4.36 Tbps. When H.265 and H.266 video compression standards are used to compress the videos, ($\frac{4.36 ~Tbps}{600}\approx$) 7.44 Gbps and ($\frac{7.44 ~Gbps}{2}\approx$) 3.72 Gbps, respectively.  

\subsubsection{\bf Practical VR with Oculus Quest 2}
\label{Sec:Quest2}
The required capacity in bits per second for a practical VR experience with Oculus Quest 2 can be derived from Eq. \ref{eq:throghput3} as follows.
\begin{align}
\label{eq:throghput2}
required~capacity = [2\times (W_{(pixels)}\times H_{(pixels)})] \\\notag\times b_{(depth)} \times frame Rate \times Compression Factor^{-1},
\end{align}
where $W_{(pixels)}$ and $H_{(pixels)}$ are the number of horizontal and vertical pixels for a single display of the HMD, respectively, $b_{(depth)}$ represent the number of bits used to represent the color of a single pixel on an image, $frame Rate$ is the number of frames (or images) displayed per second (fps), $CompressionFactor$ specifies how much a compression standard reduces the bit rate.

Quest 2 has a screen resolution of 1,832 $\times$1,920 per eye and a maximum refresh rate of 120 Hz, which supports a maximum fame rate of 120 fps. In addition, Quest 2 uses 8 bits to represent the color depth of each color channel (RGB), leading to a total of 24 bits for each pixel when no chroma subsampling is used. 
There are three options to render the video segments to be displayed on Quest 2: local rendering on the HMD, remote rendering on a server with a wired connection (using Oculus Link), and remote rendering with a wireless connection (using either the Oculus Air Link protocol or the Air Light VR [ALVR] protocol over Wi-Fi). Both ALVR and Oculus Air Link use video codecs up to HEVC (H.265), which has a maximum compression ratio of 600:1 as reported in \cite{Elbamby2018}. 

Based on these specifications and numbers, the required capacity, computed by using Eq. \ref{eq:throghput2}, is approximately equal to $\frac{2\times1,832\times1,920\times24\times120}{600}\approx32.2$ Mbps, which is far less than the human eye-like experience.

\begin{figure}[htpb!] 
    \centering 
\includegraphics[width=\linewidth]{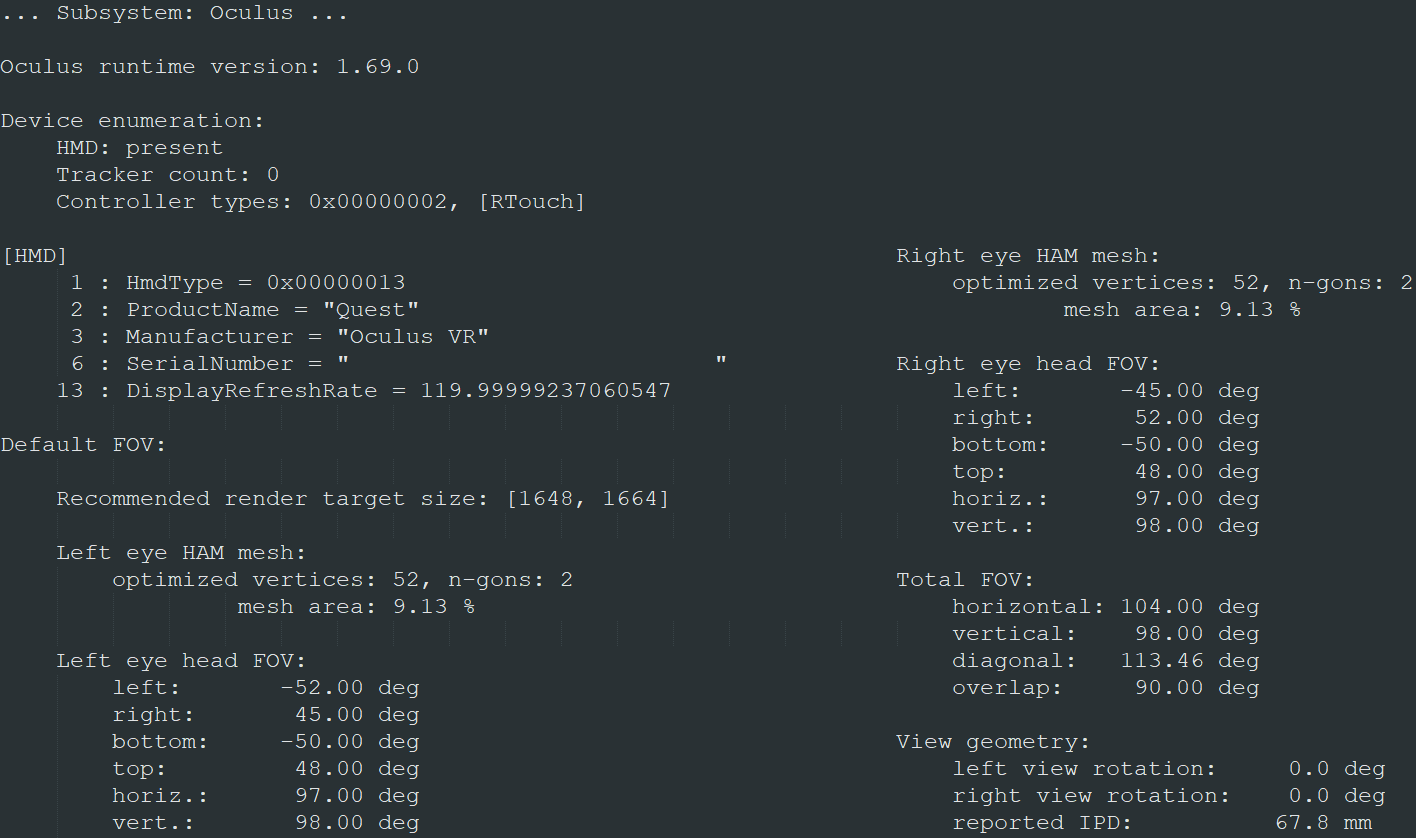} 
    \caption{HMDQ outputs of $FoV_{(h)}$, $FoV_{(v)}$, and the recommended render target size per eye for Quest 2 with 120 Hz refresh rate} 
    \label{fig:Quest2_FoV_Numbers} 
\end{figure}

\begin{figure}[htpb!] 
    \centering 
 \includegraphics[width=\linewidth]{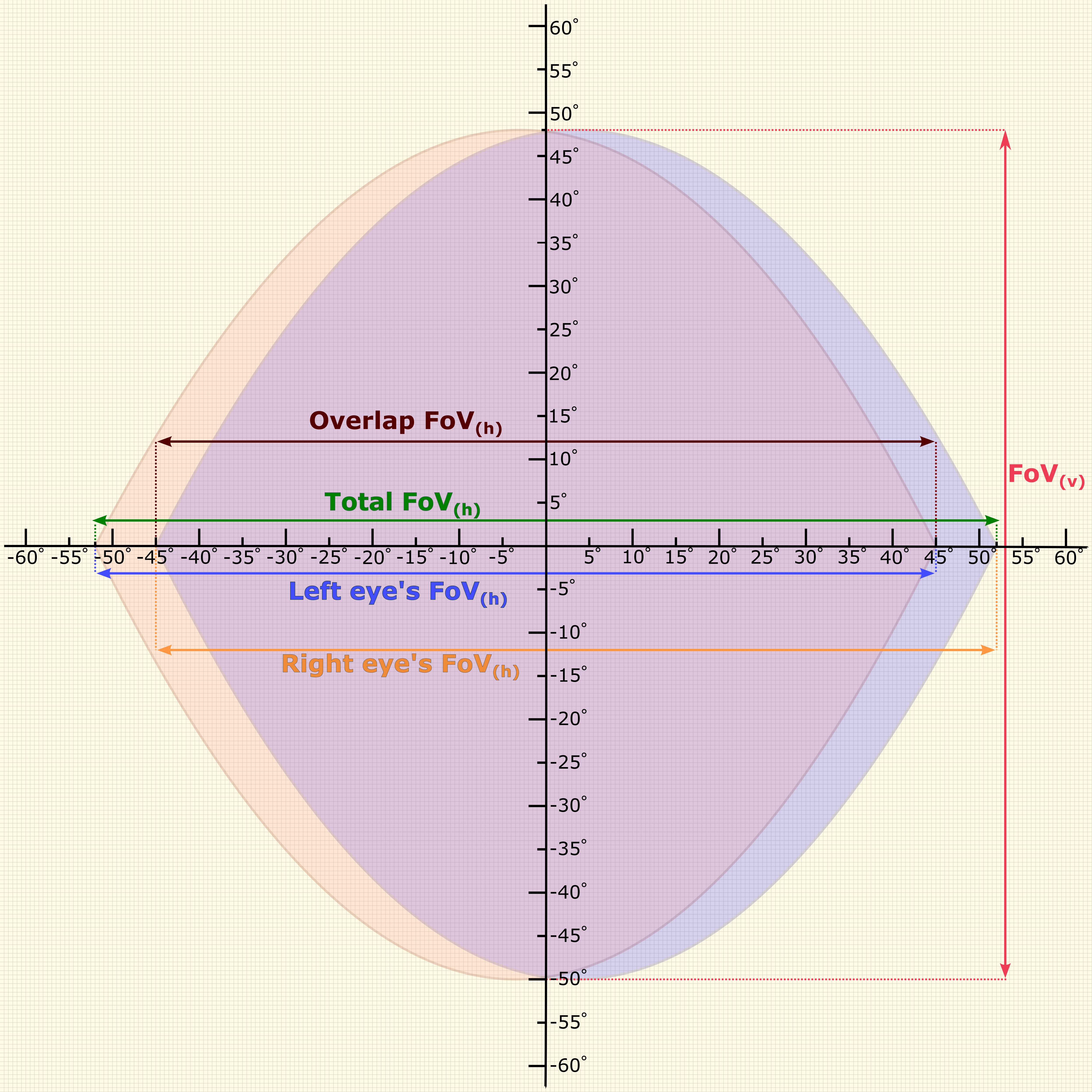} %

    \caption{$FoV_{(h)}$ and $FoV_{(v)}$ of Quest 2 measured by HMDQ tools} 
    \label{fig:Quest2_FoV_graph_V1} 
\end{figure}

This required bit rate computed is based on the specifications of Quest 2; however, the available Quest 2 specifications are missing valuable information such as the $FoV$ and $ppd$. Therefore, we have used the HMDQ\footnote{https://github.com/risa2000/hmdq} open source tools to gather Quest 2's missing information and find the recommended render target size per eye for each of its available refresh rate (72 Hz, 80 Hz, 90 Hz, and 120 Hz). By changing the refresh rate of Quest 2, the recommended highest possible resolution is changed accordingly as reported by HMDQ. For instance, Fig. \ref{fig:Quest2_FoV_Numbers} shows the measurements of Quest 2 when the refresh rate is set to 120 Hz, and Fig. \ref{fig:Quest2_FoV_graph_V1} depicts the $FoV$ of Quest 2. 
As seen in these two figures, the $FoV_{(h)}$ of both eyes is 104\degree\ with an overlap of 90\degree\ which makes the $FoV_{(h)}$ of each eye equals to 97\degree. Moreover, the $FoV_{(v)}$ of each eye is 98\degree, and the recommended render target size for each eye is 1648 $\times$ 1664. Based on the information, the quality of the image, using Eq. \ref{eq:ppd_h}, is ($\frac{1648}{97}\approx$) 16.99 $ppd$ and the required bit rate, using Eq. \ref{eq:throghput2}, is approximately ($\frac{2\times1648\times1664\times24\times120}{60}\approx$) 25.11 Mbps.

This computed bit rate is required to only transmit the viewport with the highest supported video quality by Quest 2. As mentioned before, although transmitting only the viewpor reduces the required bit rate, it needs an accurate prediction of user's head movements to fetch and transmit the appropriate viewport. Transmitting only the viewport with the  user's movement prediction (i.e., upcoming viewport prediction) is the approach used by Quest 2. To show the benefit of this approach in terms of bit rate reduction, we consider an alternative approach where the whole 360-degree is transmitted and the HMD display only the current visible viewport. To compute the required bit rate to transmit the 360-degree video, the resolution of the video needs to be computed so that the resolution of the displayed viewport meets the recommended render target size given the used refresh rate. To do so, the following equation\footnote{The ratio in the equation shows the relationship between the horizontal resolution (in terms of the number of pixels) of a single eye's viewport and the horizontal resolution of the 360-degree video by knowing the $FoV_{(h)}$ of both the video and the viewport. Similarly, the ratio that show relationship between the vertical resolution of both the video and the viewport could be derived by knowing their respective $FoV_{(v)}$.} is used to compute the resolution of either the 360-degree video or the viewport of a single eye by knowing the value of one of them and their respective $FoV_{(h)}$,
\begin{equation}
\label{eq:fullvspartRatio}
\frac{FW_{(pixels)}}{VW_{(pixels)}} = \frac{F\_FoV_{(h)}}{V\_FoV_{(h)}}, 
\end{equation}
where $FW_{(pixels)}$ is the number of horizontal pixels of the 360-degree video, $VW_{(pixels)}$ is the number of horizontal pixels for the viewport of a single eye, $F\_FoV_{(h)}$ is the horizontal field of view for the video\footnote{360-degree videos have a horizontal field of view of 360\degree and a vertical field of view of 180\degree}, and $V\_FoV_{(h)}$ is the horizontal field of view for the viewport of a single eye.

Therefore, using on the information presented in Fig. \ref{fig:Quest2_FoV_Numbers} (for Quest 2 with refresh rate of 120 Hz) and the ratio in Eq. \ref{eq:fullvspartRatio}, to get a single eye's viewport with a horizontal resolution of 1648 and a single eye's $FoV_{(h)}$ of 97\degree, the horizontal resolution of the 360-degree video has to be ($\frac{360\degree\times1648}{97\degree}\approx$) 6116 pixels. Similarly, the vertical resolution of the 360-degree video has to be ($\frac{180\degree\times1664}{98\degree}\approx$) 3056 pixels, which make the required bit rate to transmit the 360-degree video equal ($\frac{6116\times3056\times24\times120}{600}\approx$) 85.56 Mbps. The values corresponding to other Quest 2's refresh rates are presented in Table \ref{Table:Quest2Information}, and the FoVs of other HMDs (collected by HMDQ tools) could be find in \cite{Musil2022HMDDatabase}.

\begin{table*}[ht]
\centering
\begin{tabular}{||l||*{5}{c|}|}\hline\hline
\thead{\textbf{Refresh} \\ \textbf{Rate}} & \thead{\textbf{Recommended render size} \\  \textbf{target by single eye} \\\textbf{(the viewport only)}} & \thead{\textbf{The corresponding} \\ \textbf{resolution of} \\ \textbf{360-degree video}} & \thead{\textbf{\em ppd}} & \thead{\textbf{bit rate to} \\ \textbf{transmit}  \\ \textbf{only viewport} } & \thead{\textbf{bit rate to} \\ \textbf{transmit} \\ \textbf{360-degree video}}\\
      \hline\hline
\hfil\textbf{72  Hz} & 1824$\times$1840 & 6770$\times$3380 & 18.8 & 18.44 Mbps & 62.85 Mbps\\\hline
\hfil\textbf{80  Hz} & 1744$\times$1760 & 6472$\times$3232 & 17.98 & 18.73 Mbps & 63.84 Mbps\\\hline
\hfil\textbf{90  Hz} & 1648$\times$1664 & 6116$\times$3056 & 16.99 & 18.83 Mbps & 64.17 Mbps\\\hline
\hfil\textbf{120 Hz} & 1648$\times$1664 & 6116$\times$3056 & 16.99 & 25.11 Mbps & 85.56 Mbps\\\hline \hline
\end{tabular}
\captionof{table}{Summary of Quest 2 measurements for different refresh rates}
\label{Table:Quest2Information}
\end{table*}

\subsubsection{\bf Required capacity for volumetric videos}
\label{Sec:CapacityVolumetric}
The previous two derivations (human eye-like and Quest 2) are for 360-degree videos, but can be easily applied to volumetric videos with minor modifications. 

In volumetric videos with PtCl representation, a frame is represented by voxels instead of pixels as discussed in Section \ref{Sec:VolumetricVideos}. In each voxel, not only the color depth ($c_{(depth)}$) is stored, but also the position values ($p_{(depth)}$). As discussed in Section \ref{Sec:VolumetricVideos}, for each voxel, each value of the position ($x$, $y$, and $z$) is represented by 2 bytes (16 bits) which leads to a total of 6 bytes (48 bits) for all the three-position values ($p_{(depth)}=$ 48 bits). Moreover, in Eqs. \ref{eq:throghput3} and \ref{eq:throghput2}, the expression within the square brackets represents the total number of pixels at each frame for both eyes (or both displays in HMDs) and $b_{(depth)}$ represents the size of each pixel in bits which equivalent to the color depth of each pixel ($b_{(depth)}=c_{(depth)}$). 

To use these two equations to compute the required capacity for volumetric videos, we need to know the number of voxels representing each frame and the size of each voxel in bits which is equivalent to the color depth and the position depth ($b_{(depth)}=c_{(depth)}+p_{(depth)}$). Therefore, the required capacity to transmit a volumetric video (without compression) with a frame rate of 90 fps, a color depth of 24 bits (8 bits for each RGB channel), a position depth of 48 bits (16 bits for each dimension [$x$, $y$, and $z$]), and a frame quality of 50,360 voxels for each frame is equal to (50,360 voxels $\times$ [24 bits $+$ 48 bits] $\times$ 90 fps) $\approx$ $311.22$ Mbps.

\subsubsection{\bf Strong-interaction VR services}
\label{Sec:StrongInteractionVR}
\hfill \break
\phantom{2x} {\bf({\em i}) Video Compression}

In \cite{HuaweiiLab2017}, Huawei differentiates between strong-interaction and weak-interaction VR services when computing the required capacity. In strong-interaction VR services, real-time rendering, based on FoV, is required, and only the FoV is transmitted. Moreover, in the context of video compression, there are three types of video frames (pictures) clustered in groups of pictures (GOP), including intra-coded (I) frames (key frames), predicted (P) frames (forward predicted frames), and bi-directional predicted (B) frames. I-frames and P-frames are called reference frames whose information is used by (succeeding) P-frames and B-frames to predict and restore the image. In addition, P-frame and B-frame are called inter frames, which are the frames that rely on other neighboring frames to restore the image. Moreover, I-frame is self-contained enclosing the information of the image and the inter-frame encoding, which is used to decode (restore) the image. P-frame contains only the difference between itself and the closet preceding reference frames. In contrast to I-frame, the image in P-frame cannot be restored without information from previous frames (i.e., it is not a self-contained frame). The difference between a P-frame and its closet preceding I-frame or P-frame is usually small, especially in low-motion scenes, and such difference represents the movement in the scene. Thus, to save space, P-frame does not duplicate the information that could be found in the previous reference frames. Lastly, B-frame is encoded from an interpolation of the closet preceding {\em and} succeeding reference frames. Out of these three types, I-frame uses more space than the other two types, and B-frame used the least space.

\begin{figure}[htpb!] 
    \centering  
\includegraphics[width=\linewidth]{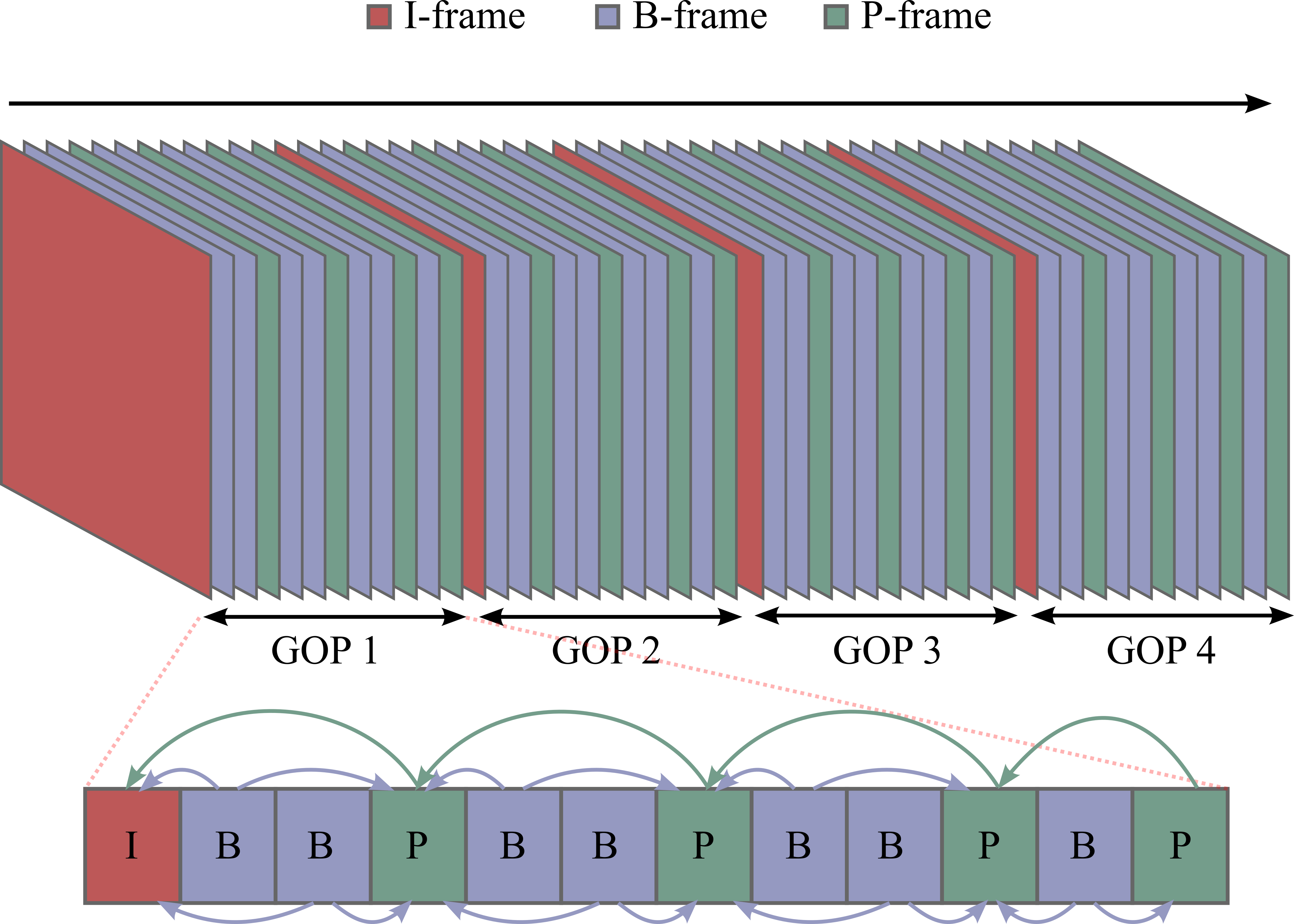} 
    \caption{Video compression's different frame types (I-frame, P-frame, and B-frame)} 
    \label{fig:Iframe} 
\end{figure}

The top illustration of Fig. \ref{fig:Iframe} depicts the different frame types used in video compression grouped in GOPs, and the bottom illustration shows the most common GOP structure (with 12 frames). The arrows in the figure indicate the reference frames for each P-frame and B-frame, respectively. In this figure, GOPs are closed groups because all the references in each group are within that group. However, in some structures, it is possible to have a P-frame or B-frame referencing a reference frame in a different GOP, and in this case, GOPs are open groups. As seen in Fig. \ref{fig:Iframe}, the length of each GOP is measured by the number of frames (e.g., 12 frames), and in each GOP, there are only one I-frame and multiple inter frames. Alternatively, the size of GOP could be measured in time. 

There is a trade-off between compression and video quality, and the size of GOP controls this trade-off. The longer the GOP, the better (in terms of reduced size) the compression, and hence, the smaller the space (or bit rate in case of streaming) needed. However, the quality of the video will be negatively affected. For instance, having GOPs with the size of one frame (i.e., all the video frames are I-frames) yields the best video quality at the cost of storage size (or bit rate in case of streaming). In contrast, it is possible theoretically to have a video with only a single I-frame and a long list of inter frames, which yields the best compression performance (reduction in storage or bit rate). However, there are two drawbacks of having only one I-frame. First, each time a specific moment in the video is sought, the encoder needs to trace back to the beginning of the video (closest I-frame) to rebuild the current frame. Second, when a frame corruption occurs (e.g., while streaming), it will be impossible to recover from this corruption since P-frames will be rebuilt on top of the corrupted frame (which means, an I-frame is required to recover from frame corruption). 

Out of the three frames mentioned above, strong-interaction VR service streams consist of only I-frames and P-frames since there are no succeeding reference frames. Therefore, there is only one I-frame for each GOP in strong-interaction VR service streams. Moreover, the number of P-frames in these streams is equal to (size of GOP $-$ 1) when the GOP size is measured in the number of frames that compose it. On the other hand, if the size of the GOP is measured in time of the duration it lasts, the number of P-frames will be equal to (GOP size in seconds $\times$ frame rate per second) $-$ 1. Based on this information and according to \cite{HuaweiiLab2017}, the required capacity in bits per second could be calculated using the following formula:
\begin{align}
\label{eq:throghput4}
required~capacity = (Size_{(iframe)} \times NB_{(iframe)} + \\\notag Size_{(pframe)} \times NB_{(pframe)}) \times (1 + redunInfo) \times \\\notag GOP Time^{-1},
\end{align}
where $Size_{(iframe)}$ and $Size_{(pframe)}$ are the size of I-frame and P-frame in bits, respectively, $NB_{(iframe)}$ and $NB_{(pframe)}$ are the number of I-frames and P-frames in each GOP, respectively, $redunInfo$\footnote{The estimated percentage of the redundancy information is 10\%\cite{HuaweiiLab2017}, which makes {\em redunInfo}'s value equals to 0.1.} is the redundancy information used for error detection and correction, and $GOP Time$\footnote{Based on \cite{HuaweiiLab2017}, the average value of GOPTime is 2 seconds.} is time duration for GOP.

\phantom{2x} {\bf({\em ii}) Timewarping (Reprojection)}

Timewarping or reprojection in VR is a technique to reduce the latency and increase or maintain the frame rate. The primary goal of timewarping is to warp the recently rendered image to make it synced with the latest captured user tracking information (TI) just before displaying the image on the HMD. A game engine, with a maximum frame rate of 50 fps, produces and sends a video frame every 20 {\em ms} based on the user TI. This results in a periodic capture-render-display operation with a cycle length of 20 {\em ms}. At the beginning of the cycle, the HMD collects the user TI, and, based on that, the game engine renders the appropriate video frame. At the end of the cycle (i.e., at time a refresh signal is received), the HMD displays the rendered video frame at the screen(s). Usually, the capture-render operation is finished early in the cycle, but the rendered video frame will not be displayed until the end of the cycle when the refresh signal is received. Therefore, the displayed video frame is based on a 20-{\em ms}-old user TI. 

One way to reduce the latency and receive a frame based on a much recent TI is by postponing the capture-render operation and making it to finish just before the refresh time of the display. However, the amount of time required for frame rendering varies from one frame to another based on the scene's complexity, and an inaccurate estimation of this time may cause the frame to be dropped due to missing the refresh time of the display, especially for intricate scenes. Another way to display a frame based on a much recent TI is to use the timewarping technique, in which, similar to the original process, the HMD captures the TI at the beginning of the cycle and renders the appropriate video frame. However, just before the display refresh time, the display recaptures a new TI, and the HMD warps the rendered video frame and changes the viewing angle to reflect this new TI using mathematical calculation. 

\begin{figure*}[htpb!] 
    \centering  
\includegraphics[width=0.9\linewidth]{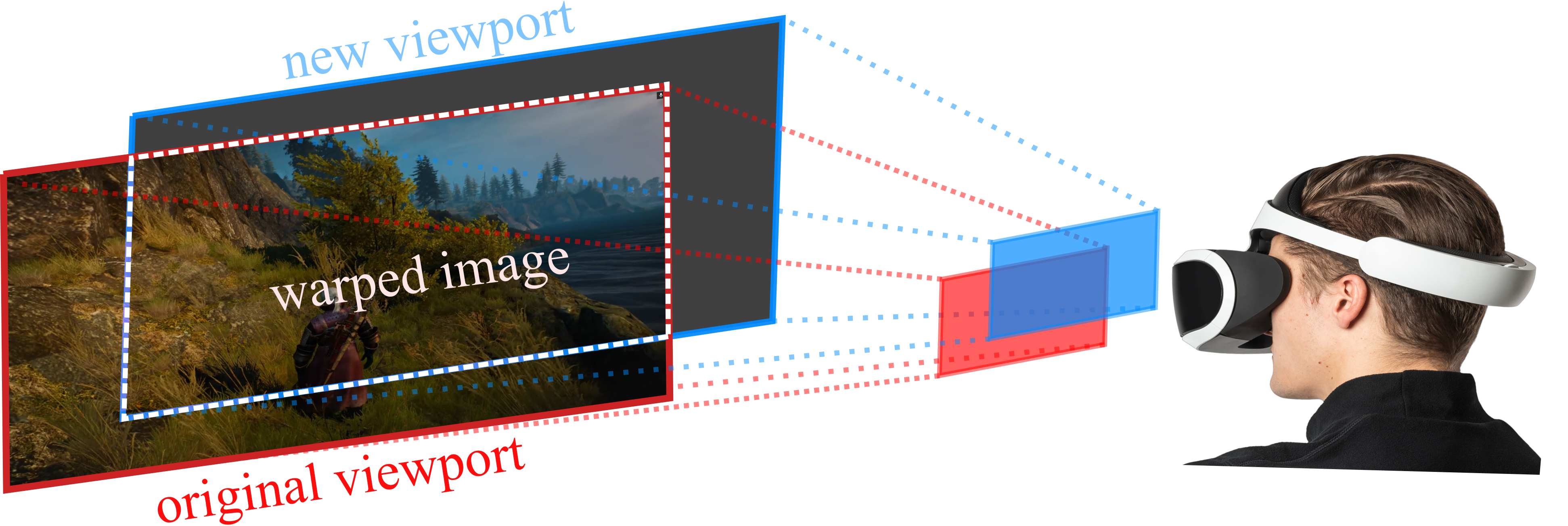} 
    \caption{Timewarping in VR} 
    \label{fig:timewarping} 
\end{figure*}

In addition to reducing the latency, the timewarping technique could help increasing the frame rate. For instance, a display with a refresh rate of 72 Hz can present frames up to 72 fps. However, with a game engine that produces frames at a rate of 50 fps, the display will not be able to achieve its potential. Instead, it will be limited to the frame rate of the game engine. In this case, timewarping could be used to create more frames by warping the most recent frame based on the most recent TI. Fig. \ref{fig:timewarping} depicts the timewraping process in VR. The viewer in this figure was seeing the red viewport; then, they changed the head orientation to see the blue viewport. Therefore, timewarping warps the rendered red viewport and creates a warped image within the dotted white area. Moreover, as seen in the figure, the warped image is smaller than the viewport, and the rest of the blue viewport (the new viewport) is black. 

To eliminate this black area, the rendered frames should have dimensions larger than the FoV of the HMD as considered in Eq. \ref{eq:NBofPixels}. For example, if $FoV_{(h)}$ and $FoV_{(v)}$ of an HMD is 120$\degree$ and 90$\degree$, respectively, the rendered video frame should correspond to $FoV_{(h)}$ and $FoV_{(v)}$ of 132$\degree$ and 102$\degree$, respectively, with an additional of 12$\degree$ for both horizontal and vertical FoVs. Moreover, to support timewarping, additional depth of field information is required to adjust the viewing angle to reflect the most recent TI as considered in Eq. \ref{eq:frameSize}. 

Based on the aforementioned information and according to \cite{HuaweiiLab2017}, $Size_{(iframe)}$ and $Size_{(pframe)}$ in Eq. \ref{eq:throghput4} could be computed as follow.
\begin{align}
\label{eq:frameSize}
Frame~size = NB_{(pixels)} \times b_{(depth)} \times \\ (1+\notag DOF_{(info)}) \times CompressionFactor^{-1},
\end{align}
where $NB_{(pixels)}$ is the total number of pixels in the frame, $b_{(depth)}\footnote{In \cite{HuaweiiLab2017}, they assume that a chroma subsampling of (4:2:0) ratio is used, which makes the $b_{(depth)}$ of 8-bit, 10-bit, 12-bit, and 16-bit color equals to 12, 15, 18, and 24 bpp, respectively.}$ is the bit depth as the number of bits per pixel,  $DOF_{(info)}$\footnote{It is the percentage of additional information of the depth of field used in timewrapping. Based on \cite{HuaweiiLab2017}, timewrapping requires 15\% additional information, which makes $DOF_{(info)}$ = 0.15.} is the percentage of additional information of the depth of field, and $CompressionFactor$\footnote{This value is different from one video to another based on the frame rate and the resolution. Moreover, the compression factor of I-frames is usually smaller than the compression factor of P-frames} is the compression factor of the encoding standard. Moreover, $NB_{(pixels)}$ is computed as follow \cite{HuaweiiLab2017}.
\begin{align}
\label{eq:NBofPixels}
NB_{(pixels)} = 2\times (W_{(pixels)}\times H_{(pixels)})\times \\ (1+\notag\frac{FoV_{(h)}\prime}{FoV_{(h)}})\times(1+\frac{FoV_{(v)}\prime}{FoV_{(v)}})\times \\ (1+\notag FoV_{(pictures)})^2,
\end{align}
where $W_{(pixels)}$ and $H_{(pixels)}$ are the number of horizontal and vertical pixels for a single display of the HMD, $FoV_{(h)}$ and $FoV_{(v)}$ are the horizontal and vertical Field of Views, $FoV_{(h)}\prime$ and $FoV_{(v)}\prime\footnote{The extra perspective degrees, according to Huawei, are 12 degrees vertically (6 degrees top and 6 degrees bottom) and 12 degrees horizontally (6 degrees left and 6 degrees right)}$ are the extra perspective of the horizontal and vertical Field of Views degrees, and $FoV_{(picture)}$\footnote{It is about 10\% \cite{HuaweiiLab2017}} is the percentage of the extra  picture in $FoV$.

\phantom{2x} {\bf({\em iii}) An Example of Strong-interaction VR Service}

In \cite{HuaweiiLab2017}, Huawei envisioned the evolution of VR in three phases, each of which has multiple parameters. Out of these phases, we are living in the era of the comfortable-experience phase. In this phase, a single eye resolution of 1920 $\times$ 1920 is expected with a color depth of 8 bpc, which results in 12 bpp if a chroma subsampling of (4:2:0) ratio is used. In addition, both horizontal and vertical $FoV$ are expected to be 120$\degree$, in this phase, with a frame rate of 90 fps. The used encoding standard H.265 makes the compression ratios approximately 38:1 and 165:1 for I-frame and P-frame, respectively.

To compute the typical bit rate using the aforementioned parameters, Eqs. \ref{eq:throghput4}, \ref{eq:frameSize}, and \ref{eq:NBofPixels} are used. Using Eq. \ref{eq:NBofPixels}, the number of pixels of a frame is equal to $2\times(1920\times1920)\times(1+\frac{12}{120})\times(1+\frac{12}{120})\times(1+0.1)^2\approx10,794,516$ pixels per frame. Moreover, from Eq. \ref{eq:frameSize}, the sizes of I-frame and P-frame are equal to $10794516\times12\times(1+.15)\times38^{-1}\approx3,920,114$ bits and $10794516\times12\times(1+.15)\times165^{-1}\approx902,814$ bits, respectively. Therefore, the typical bit rate, according to Eq. \ref{eq:throghput4}, equals ($3920114\times1+902814\times[(2\times90)-1])\times(1+0.1)\times2^{-1} \approx 91,038,101$ bps $\approx 91$ Mbps.

\subsection{\bf Related work}
\label{sec:capacity_relatedwork}
Many papers have discussed and analyzed the required throughput for immersive XR experiences under different parameters. In \cite{Elbamby2018}, the authors consider a human-eye-like VR view with horizontal and vertical FoV equals 150\degree\ and 120\degree, respectively. Their analysis considers a vision acuity of exactly 20/20 (translated to 60 $ppd$) and a frame rate of 120 fps. Thus, the required bit rate is 1 Gbps when the used compression protocol is H.256, and the bit depth is 36 bits for each pixel. For an `entry-level VR' (with a resolution of 1K or 2K), the required throughput could reach 100 Mbps \cite{Elbamby2018}. 

The authors of \cite{Bastug2017} use similar configurations to estimate the minimum required throughput to achieve a human eye-like VR experience. However, in their estimation, a pixel density equivalent to 200 $ppd$ and a fame rate equal to 60 fps is used. Therefore, the required throughput is at least 5.2 Gbps \cite{Bastug2017}. For a 4K VR experience with 30 fps and 24 bits per pixel ($bpp$) bit depth, the minimum throughput is 300 Mbps if a compression standard with a compression ratio of 300:1 is used \cite{Bastug2017}. 

In \cite{ABIResearchQualcomm2017}, the authors define different types of XR-related experiences, including monoscopic 4K 360-degree videos, stereoscopic HDR 360-degree videos with very high resolutions ($\geq$ 8K) and high frame rates ($\geq$ 90 fps), and 6DoF (volumetric) videos or Point Cloud (PtCl). The first two types of experience allow head rotational movements only, while the last type allows both body translational and head rotational movements. The required bandwidth of the monoscopic 4K videos, the  stereoscopic HDR videos, and the 6DoF videos are 10-50 Mbps, 50-200 Mbps, and 200-1000 Mbps, respectively. 

In 2016, Huawei
envisioned the evolution of VR experiences and categorized them into four stages, namely pre-VR, entry-level VR, advanced VR, and ultimate VR \cite{HuaweiTechnologies2016}. For each stage, a time frame had been specified for its commercialization together with the user's {\em continuous experience duration} and several other parameters that identify the stage. Specifically, the pre-VR stage started in 2016, while the entry-level VR stage started in 2018. In these two stages, users have a continuous experience duration of fewer than 20 minutes. We are now in the advanced VR stage with the user's continuous experience duration ranging between 20 and 60 minutes. The ultimate VR stage is expected in the coming years with the user's continuous experience duration exceeding one hour.

The other user experience parameters in each phase as defined in \cite{HuaweiTechnologies2016} can be summarized in the tuple [single-eye resolution in pixels$\times$pixels, frame rate in fps for weak interaction/strong interaction, color depth in bpc, field of view of a single eye in degree$\times$degree, the used encoding standard]. The pre-VR stage, having parameters [1080$\times$1200 pixels, 30/90 fps, 8 bpc, 100\degree$\times$111\degree, H.264], has an estimated required bit rates of 16 Mbps and 18 Mbps for the weak-interaction and strong-interaction VR services, respectively. The entry-level VR stage with parameters [1920$\times$1920 pixels, 30/90 fps, 8 bpc, 110\degree $\times$110\degree, H.265] needs bit rates of approximately 50 Mbps and 80 Mbps to transmit 2D and 3D full-view videos, respectively, for the weak-interaction VR services. To transfer only the $FoV$ for the weak-interaction VR services in this phase, bit rates of 26 Mbps and 42 Mbps are required for 2D and 3D videos, respectively. For the strong-interaction VR services, bit rates of 40 Mbps and 60 Mbps are required to transmit the $FoV$ for 2D and 3D videos, respectively. The advanced VR and ultimate VR phases have the parameters of [3840$\times$3840pixels, 60/120 fps, 10 bpc, 120\degree$\times$120\degree, H.265] and [7680$\times$7680 pixels, 120/200 fps, 12 bpc, 120\degree$\times$120\degree, H.266], respectively. The required bit rates to transmit the full-view videos in the weak-interaction VR services are approximately 420 Mbps and 2.94 Gbps for the advanced VR and the ultimate VR stages, respectively. To transmit only the $FoV$, these bit rates are reduced to 220 Mbps and 1.56 Gbps for the weak-interaction VR services. For the strong-interaction VR services, the required bit rates to transmit the $FoV$ for these two stages are 390 Mbps and 680 Mbps, respectively.

The authors of \cite{Hu2020} considered not only the four VR stages described by Huawei in \cite{HuaweiTechnologies2016}, namely pre-VR, entry-level VR, advanced VR, and ultimate VR, but also human precision as another stage. The parameters of user experience in each stage can be summarized as the tuple [angular view resolution in $ppd$, frame rate in fps, color depth in bpc, field of view of a single eye in degree$\times$degree]. Moreover, for each stage, the authors estimate the required bit rate in three scenarios: uncompressed, transmission with a low latency compression having a compression ratio of 20:1, and transmission with a lossy compression (such as H.264) having a compression ratio of 300:1.  The pre-VR stage, having parameters [10 $ppd$, 60 fps, 8 bpc, 100\degree$\times$100\degree], has an estimated required bit rates of 10.62 Gbps, 530 Mbps, and 35 Mbps for the first, second, and third scenarios, respectively. The entry-level VR stage with [17 $ppd$, 90 fps, 8 bpc, 110\degree $\times$110\degree] needs approximately 63.7 Gbps bit rate if no compression is used. If a low latency compression with a compression ratio of 20:1 is used, the approximate required bit rate for the entry-level VR will be reduced to become 3.18 Gbps for the full view (of 360\degree $\times$180\degree) and 796 Mbps for only the FoV. A lossy compression encoding will further reduce these requirements to 210 Mbps for the full view and 53 Mbps for the FoV. The advanced VR, human precision, and ultimate VR stages have the parameters of [32 $ppd$, 120 fps, 10 bpc, 120\degree $\times$120\degree], [60 $ppd$, 120 fps, - , 150\degree $\times$135\degree], and [64 $ppd$, 200 fps, 12, 150\degree $\times$150\degree], respectively. The required bit rates for these stages, if no compression is used, are approximately 238.89 Gbps, 1007.77 Gbps, and 1911.03 Gbps, respectively. Using a compression protocol with a compression ratio of 20:1 will reduce bit rates requirements to 11.94 Gbps, 50.39 Gbps, and 95.55 Gbps for the full view and 5.31 Gbps, 31.49 Gbps, and 66.36 Gbps for the FoV for the advanced, human precision, and ultimate stages, respectively. Lossy compression with a compression ratio of 300:1 will further reduce the required bit rates for the full view case to 796 Mbps, 3.36 Gbps, and 6.37 Gbps for the advanced, the human precision, and ultimate VR stages, respectively. It will also reduce the required bit rates for the FOV case to 354 Mbps, 2.10 Gbps, and 4.42 Gbps for each of these stages, respectively.

Simone Mangiante et al. \cite{Mangiante2017} categorized VR experiences into four stages, namely early stage, entry-level, advanced, and extreme. For each stage, a set of parameters are defined, including the resolution in the FoV area, the type of VR (monoscopic [2D] or stereoscopic [3D]), the frame rate, the channel color depth (bits per RGB channel). The first three stages are 2D (monoscopic) VR experiences, while the last one is 3D (stereoscopic) experience. The first two stages have 30 fps, while the third and the fourth stages have 60 fps and 120 fps, respectively. Moreover, in the first two stages, the channel color depth is 8 bpc (24 bpp), while in the third and the fourth stage, the color depths are 10 bpc (30 bpp) and 12 bpc (36 bpp), respectively. The resolutions of the visual fields are 1K$\times$1K, 2K$\times$2K, 4K$\times$4K, and 8K$\times$8K for the first, second, third, and fourth stage, respectively. The required throughput for each stage is 25 Mbps for the first stage, 100 Mbps for the second stage, and 400 Mbps for the third stage. The fourth stage needs 1 Gbps for smooth plays and approximately 2.35 Gbps for interactive plays.

For nowadays XR experiences, the required throughput varies based on the type of XR application. For in-vehicle services, such as the one mentioned in Fig. \ref{fig:car}(d), the required throughputs could be up to 20 Mbps \cite{Huawei2016}. On the other hand, most of the current low-resolution AR/VR streaming applications need around 25 Mbps \cite{Huawei2016}. Generally speaking, in today's standards, the required throughput for a decent VR experience is between 25 Mbps and 500 Mbps, and the required throughput for an AR experience is between 1 Mbps and 200 Mbps \cite{Adame2020}. Moreover, based on Eq. \ref{eq:throghput2}, the maximum raw data rate resulting from Oculus Rift and HTC Vive, with a maximum frame rate of up to 90 fps and a resolution of 2,160$\times$1,200, is approximately 5.6 Gbps without optimization or compression and with 8 bpc \cite{Huawei2016}. In the near future, VR is aiming to achieve a resolution of 4K and 8K UHD (3840$\times$2160 and 7680$\times$4320), which require approximate higher throughput of 17.9 and 71.7 Gbps, respectively \cite{Huawei2016}. Moreover, since 4K 360-degree VR videos have a pixel density of approximately 10 $ppd$, which is equivalent to regular videos with a resolution of 240p, the future VR has to have a resolution of at least 8K or more to achieve a pleasant VR experience \cite{Huawei2016}. Using Eq. \ref{eq:fullvspartRatio}, a 360-degree video with a resolution of 4K ($4096 \times 2160$) would have a viewport resolution (for a single eye) of 1K ($1024 \times 720$), if $FoV_{(h)}$ and $FoV_{(v)}$ (per eye) are 90\degree and 60\degree, respectively. Similarly, to achieve a viewport resolution (for a single eye) of 4K ($4096 \times 2160$) with the same $FoV_{(h)}$ and $FoV_{(v)}$, the 360-degree video needs to have a resolution of 16K ($16384 \times 6480$) \cite{Mangiante2017}.
\section{The Latency Requirement of XR} 
\label{Sec:Latency}


\begin{table*}[htbp]
\centering
\begin{tabular}{||p{4cm}||p{6cm}||p{5.5cm}||}\hline
\hline\makebox[4cm]{Classification Title}&\makebox[6cm]{Classification Value}&\makebox[5.7cm]{Papers} \\\hline\hline

\multirow{3}{*}{\makecell{The type of streamed \\ video}}&on-demand 360-degree video& \cite{Bao2016ShootingVideos, Boos2016, Chakareski2018, Corbillon2017, Fan2017FixationReality, Fu2019, Ghosh2017, Ghosh2018, Guntur2012OnLAN, Hu2019, Jiang2020, Kammachi-Sreedhar2017, Kan2019DeepStreaming, Krouka2020, Liu2018, Mavlankar2010VideoPan/Tilt/Zoom, Petrangeli2017AnVideos, Qian2016, Qian2018Flare:Devices, Sassatelli2019, Zhang2019DRL360:Learning}\\\cline{2-3}
&live 360-degree video& \cite{Aksu2018, Kimata2012MobileSystem, Liu2019, Mavlankar2009Pre-fetchingSequences}\\\cline{2-3}
&volumetric video& \cite{Liu2019a, Qian2019}\\\hline\hline

\multirow{5}{*}{\makecell{Minimization approach \\ used }}&optimization-based approach&  \cite{Al-Shuwaili2017, Bohez2013, Verbelen2013}\\\cline{2-3}
&parallel transmission and computation& \cite{Liu2018}\\\cline{2-3}
&tile-based adaptive approaches& \cite{Aksu2018, Bao2016ShootingVideos, Chakareski2018, Fan2017FixationReality, Fu2019, Ghosh2017, Ghosh2018, Guntur2012OnLAN, Jiang2020, Kan2019DeepStreaming, Kimata2012MobileSystem, Krouka2020, Mavlankar2009Pre-fetchingSequences, Mavlankar2010VideoPan/Tilt/Zoom, Petrangeli2017AnVideos, Qian2016, Qian2018Flare:Devices, Sassatelli2019, Zhang2019DRL360:Learning}\\\cline{2-3}
&projection-based adaptive approaches& \cite{Corbillon2017, Hu2019, Huawei2016, HuaweiiLab2017, Kammachi-Sreedhar2017}\\\cline{2-3}
& 
retrieval-based approaches& \cite{Boos2016,Liu2019a}\\\hline\hline

\multirow{9}{*}{\hfil \makecell{Viewport prediction \\ approach}}& heuristic algorithms and  probabilistic approaches& \cite{Chakareski2018, Ghosh2017, Ghosh2018, Hu2019}\\\cline{2-3}
&linear regression (LR)& \cite{Bao2016ShootingVideos, Qian2018Flare:Devices}\\\cline{2-3}
&ridge regression (RR)& \cite{Qian2018Flare:Devices}\\\cline{2-3}
&support vector regression (SVR)& \cite{Qian2018Flare:Devices}\\\cline{2-3}
&neural network (NN)& \cite{Aksu2018, Bao2016ShootingVideos}\\\cline{2-3}
& recurrent neural network (RNN) [e.g., long short term memory (LSTM)]& \cite{Fan2017FixationReality}\\\cline{2-3}
&reinforcement learning (RL)& \cite{Jiang2020}\\\cline{2-3}
&deep reinforcement learning (DRL)& \cite{Kan2019DeepStreaming, Krouka2020, Sassatelli2019, Zhang2019DRL360:Learning}\\\cline{2-3}
& sequential deep reinforcement learning (SDRL)& \cite{Fu2019}\\\hline\hline

\multirow{3}{*}{\hfil \makecell{Input information for \\ viewport prediction}}& the current and past user head rotation& \cite{Bao2016ShootingVideos, Chakareski2018, Qian2018Flare:Devices}\\\cline{2-3}
& the current and past user head rotation in addition to content-related information & \cite{Fan2017FixationReality, Mavlankar2009Pre-fetchingSequences, Mavlankar2010VideoPan/Tilt/Zoom, Nguyen2018YourPrediction}\\\cline{2-3}
& crowdsource information (from multiple users) & \cite{Ghosh2018, Guntur2012OnLAN, Liu2019}\\\hline\hline

\multirow{2}{*}{\hfil The transmitted parts}& transmit $only$ the viewport in the highest possible resolution & \cite{Bao2016ShootingVideos, Fan2017FixationReality, Hu2019, Liu2019, Qian2016, Qian2018Flare:Devices}\\\cline{2-3}
& transmit the viewport in the highest possible resolution and the rest of video in lower resolution & \cite{Aksu2018, Chakareski2018, Corbillon2017, Fu2019, Ghosh2017, Ghosh2018, Jiang2020, Kammachi-Sreedhar2017, Kan2019DeepStreaming, Krouka2020, Petrangeli2017AnVideos, Sassatelli2019, Zhang2019DRL360:Learning}\\\hline\hline

\multirow{2}{*}{\hfil The used technology}& mobile devices for display & \cite{Liu2019, Qian2018Flare:Devices}\\\cline{2-3}
& HTTP and HTTP/2 for transmission (using the Dynamic Adaptive Streaming over HTTP (DASH) approach) & \cite{Liu2019, Petrangeli2017AnVideos, Qian2018Flare:Devices}\\\hline\hline

\end{tabular}
\captionof{table}{Classifications of existing XR streaming research}
\label{Table:currentResearch}
\end{table*}


The other critical factor for any immersive XR experience is the motion-to-photon (MTP) latency, which is the time required to display a scene (video segment) corresponded to HMD's movement (translational, rotational, or both) and other interactions and activities. For XR over {\em wireless} edge computing (the scenario we consider in our paper), MTP latency is the round trip delay from the HMD to the edge server and back to the HMD. High and unstable MTP latency leads to low user experience and, for VR in particular, motion sickness. In addition to processing delay, XR over edge introduces the communication latency, which is higher and time-varying in wireless compared to the wired counterpart (via an USB or HDMI cable to a local server).

This section discusses the major delay components of MTP and summaries the maximum acceptable delay for each component to facilitate pleasant immersive XR experiences. The section will first review the multiplayer exercise games and discuss which game data delay (on the player's side or on the opponents' sides) has the highest impact on this type of VR applications \cite{Kojic2019}. It then thoroughly analyzes the main delay components of MTP, including the delays introduced by the communications between the HMD and the edge on one side and between the edge and the cloud on the other. Furthermore, it will discuss some of the optimizations that have been performed to reduce MTP latency. Finally, it will describe various types and classifications of XR experiences and show the required delay for each type to achieve satisfying immersive XR experiences.

\subsection{\bf Multiplayer exercise games and their delay analysis}

One common type of VR games is the multiplayer exercise game or {\em exergame} for short, where games integrate with physical exercises (i.e., movements) with multiplayer competition. There are two types of data in exergames that need to be processed either locally or on the edge. The two types of data are the {\em scene data} required to render the scenes, such as the user's head orientation, and the {\em game data}, such as the speed of rowing, required to render the virtual players and other dynamic objects in the scenes \cite{Kojic2019}. Typically, the scene data has more stringent delay and reliability requirements than the game data which is more delay tolerant. When rendering is processed on the edge devices, these two types of data could be sent to the edge device through either the same communication channel or different channels.

In \cite{Kojic2019}, the authors analyze the effect of different delays on users' experience in a multiplayer exergame. In their experiments, a real user competes in a rowing race with a virtual opponent within a virtual environment seven times, each with a different delay setting. Specifically, the scene data is rendered locally based on the orientation of the HMD. Simultaneously, the game data generated by the real user is sent wirelessly from a rowing ergometer to an edge device through a smartphone. 
The experiments focused on the communication delays, experienced by the real user and simulated for the virtual opponent, respectively, that affect the \underline{\em game data} (such as the speed and the distance traveled as measured by the rowing ergometer),
and tuned the delay to represent three levels of latency: low (30 ms), medium (100 ms), and high (500 ms). These three levels result in multiple settings of delay combinations between the real user and the virtual opponent. The worst-case scenario is to have high delays for both the user and the opponent, and the best-case scenario is to have low delays for both. Using different delay settings, 23 participants (17 males and 6 females) were asked to play against a virtual opponent and answer multiple questionnaires to evaluate their VR experiences based on the overall quality of experience (QoE), the presence and flow of the opponent, and the perceived delay.

Based on the presented results, the human player's own delay has a higher impact on the QoE, flow, and presence compared with the virtual opponent's. In addition, although delay occurs on both sides, the human could not always correctly determine on which side the delay occurred. For instance, the human players rated the opponent's delay high, even though it was low, due to the fact that their own delays were high. On the other hand, the human players rated QoE as good when their own delays were low, regardless of the opponent's delay. Moreover, there was no significant difference, in terms of QoE, between the low and medium delay scenarios because there is much lesser interaction between the players in this type of games (racing games) compared to other multiplayer games, such as shooting games.

In the next subsection, we discuss the MTP latency in edge-enabled wireless XR scenarios and the main components that contribute to this latency. 

\subsection{\bf MTP latency of edge-enabled wireless XR}

MTP is the time lapse between a head or body movement and the display of its corresponding scene on the HMD. For VR,
a high MTP value causes conflicting signals to be sent to Vestibulo-ocular Reflex (VOR), a part of the cranial nerves that consist of three efferent neurons, CN-III (oculomotor), CN-IV (trochlear), and CN-VI (abdducens), which are the motor neurons controlling the muscles around the eyes \cite{Somisetty2021NeuroanatomyReflex}. The primary function of VOR is to stabilize the gaze during the head movements by using the information from the sensory signals. However, conflicts in these signals, caused by the high MTP delay, lead to motion sickness. 
In general, to avoid motion sickness in VR, there is a broad consensus that an upper bound of MTP latency between 15 ms and 20 ms would be necessary \cite{Elbamby2018, Liu2018, ABIResearchQualcomm2017, Mangiante2017, Huawei2016, Boos2016, Abrash2014,Nasrallah2019}. Others suggested even more stringent MTP latency requirements of up to 10 ms \cite{Hu2020,Adame2020}, 8 ms \cite{HuaweiiLab2017}, 7 ms \cite{Mangiante2017}, and 5 ms \cite{Hu2020} for ultimate and extreme VR phases/stages and/or strong-interaction VR services. On the other hand, to enable a pleasant AR experience, \cite{Adame2020} suggested a maximum MTP latency between 1 ms and 50 ms for different AR applications, and for infotainment AR, \cite{Nasrallah2019} argued for the maximum MTP latency between 7 ms and 20 ms.

\begin{figure}[htpb!] 
    \centering 
\includegraphics[width=\linewidth]{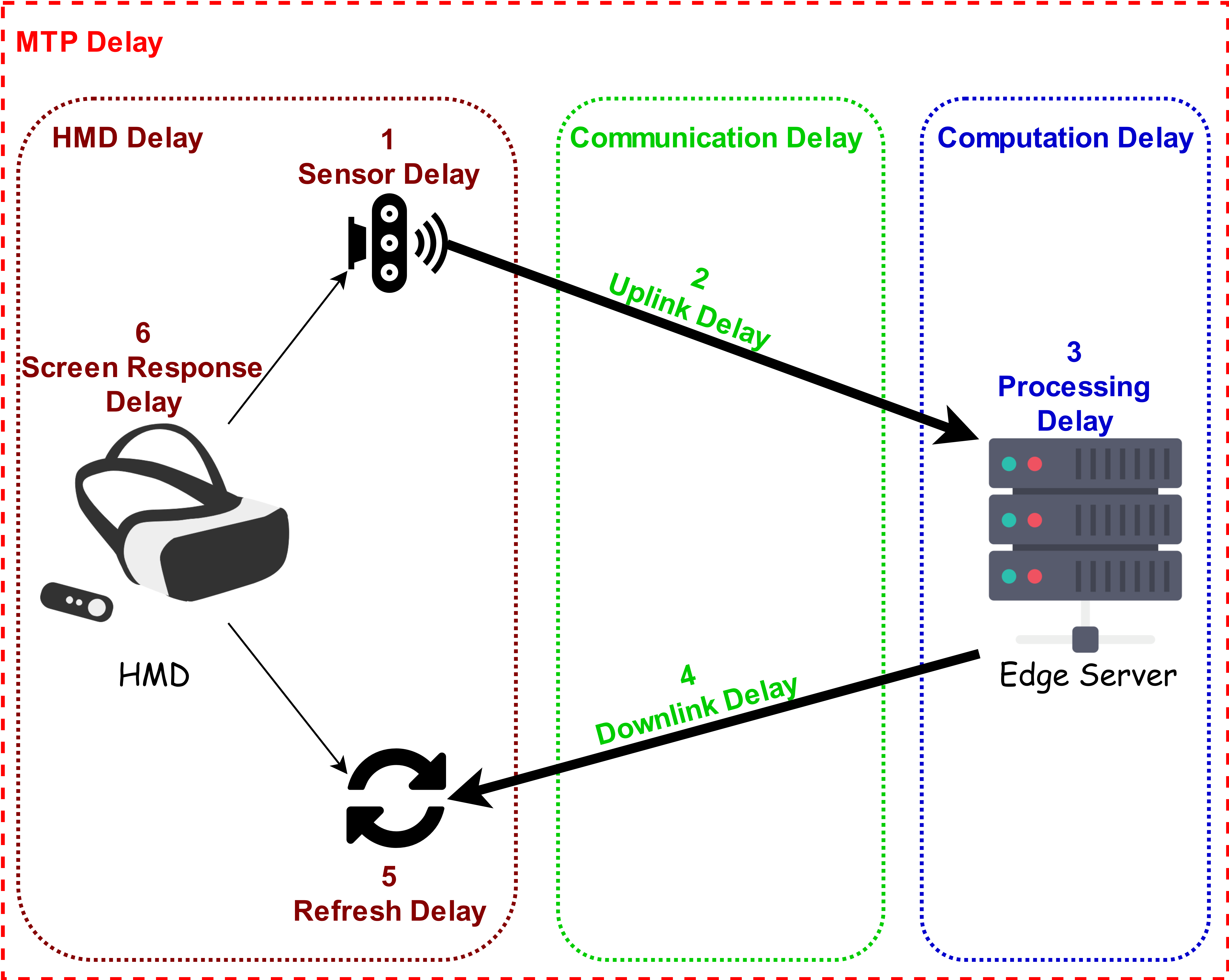} 
    \caption{Delay components of MTP} 
    \label{fig:delayComponents} 
\end{figure}

Fig. \ref{fig:delayComponents} depicts the six main delay components of MTP described in \cite{Huawei2016}, divided into three categories of HMD-related delay, communication delay, and computation delay. As seen in the figure, the HMD-related delay consists of the sensing delay, the screen response delay, and the refresh delay. On the other hand, the communication delay (network round-trip-time [RTT\footnote{Note that the term RTT was used in some papers as a synonym for MTP by also including HMD and computation delays.}]) composes of the time of sending the scene and game data from the HMD to the edge device (the uplink delay) and the time of receiving the corresponding rendered video segments by the HMD from the edge (the downlink delay). Moreover, for the communication delay, the network RTT includes queuing, transmission, and propagation delay. Finally, the computation delay is incurred by processing the data received by the edge from the HMD through the uplink. This process includes complex tasks such as 3D image processing, 3D dynamic audio tracing, color and lens deformity correction, frame rendering, and video encoding. For AR specifically, scene reconstruction using 3D models will also be needed.

Out of the aforementioned delay components,
the sensor sampling latency is extremely small (less than 1 ms) \cite{Huawei2016, Elbamby2018}. Moreover, the screen response delay (pixel delay), which is the required time to convert a pixel from one color state to another, is at most 2 ms \cite{Huawei2016}. In addition, the refresh delay of a screen with 120 Hz is at most $\frac{1}{120}\times1000\approx 8.3$ ms. The previous values together add up to 11 ms to form the HMD delay depicted in Fig. \ref{fig:delayComponents}
In addition, although the display delay, which includes both the refresh delay and the screen response delay, using the current technology is in the range of 10--15 ms \cite{Elbamby2018}, it is expected to drop to approximately 5 ms with the help of dynamic refresh \cite{Mangiante2017}. Given the upper bound of 20 ms for MTP, about 14 ms are left for both the communication and computation delays. Moreover, the edge server needs 2 ms to process the incoming data from the HMD and produce the appropriate video segments to send back to the HMD \cite{Huawei2016}. Therefore, the communication delay (both uplink and downlink delay) has to be in the range of 8 to 12 ms with an average of 10 ms. Unlike AR and MR, for VR applications, we need to transmit much more data (i.e, the rendered videos) in the downlink compared to the uplink (i.e., the user tracking information). Therefore, it is wise to dedicate a larger portion of the MTP delay budget to the downlink.


The authors of \cite{Mangiante2017} compared the VR use cases of local VR and online VR with mobile edge computing (MEC). In the case of local VR, the HMD is connected, using an HDMI or USB cable, to a server which render raw video without any compression used. In this scenario, the sensor latency is almost 1 ms, and the display latency is 10 ms, which leads to the total HMD latency of 11 ms. The required computation time for rendering on the server is in the range of 3--5 ms and the communication delay is almost 2 ms, which adds up to an overall MTP latency in the range of 15--18 ms. 

In contrast, online VR uses MEC, video compression, and dynamic refresh to render, compress, and display the video segments. Although the sensor delay will remain the same (1 ms), the display delay is expected to drop to 5 ms, owing to the use of dynamic refresh to present the decompressed video segments on the HMD. In addition, the communication latency and the computation delay are expected to be 5 ms and 7 ms, respectively, which results in an MTP latency of 18 ms.

In \cite{Liu2018}, the authors divided the MTP latency into four components, the sensing latency, the rendering latency, the streaming latency, and the display latency, as follow.
\begin{equation}
\label{eq:e2elatency}
T_{e2e} = T_{sense} + T_{render} + T_{stream} + T_{display} 
\end{equation}
where
\begin{equation}
\label{eq:streamlatency}
T_{stream} = T_{encode} + \underbrace{T_{trans}}_{\frac{frameSize}{Throughput}} + T_{decode} 
\end{equation}

$T_{e2e}$ denotes the MTP latency, which includes all the components inside the dashed red rectangle in Fig.\ref{fig:delayComponents}. $T_{sense}$ represents the time required to transmit the sensor data to the edge server, which is shown as delay components \#1 and \#2 in Fig.\ref{fig:delayComponents}. $T_{render}$ is the time required by the edge server to render video segments, which represents the majority of delay component \#3 in the figure. $T_{stream}$ is the time needed to send a rendered video segments to the HMD, which includes the time to compress a rendered frame on the edge server ($T_{encode}$), the time to transmit the compressed frame from the edge server to the HMD ($T_{trans}$), and the time to decompress the received frame in the HMD ($T_{decode}$). $T_{encode}$ represents the rest of delay component \#3 in Fig. \ref{fig:delayComponents}, while $T_{trans}$ and $T_{edcode}$ represents delay component \#4 and part of the HMD delay, respectively. Moreover, the transmission delay ($T_{trans}$) could be computed by using the formula $\frac{L}{R}$, where $L$ is the data size (the compressed frame size) and $R$ is the available bit rate. Finally, $T_{display}$ is the time needed by the HMD to display the decoded frames, i.e., the HMD delay components depicted in Fig. \ref{fig:delayComponents}. Each screen generates periodic VSync signals based on its refresh rate, and if the received decoded frame missed the current signal, it has to wait in the buffer for the next signal, which incurs the display latency. The refresh delay, shown in Fig. \ref{fig:delayComponents}, is affected by the screen's refresh rate. A screen with a refresh rate of 90 Hz has a maximum waiting time of approximately 11.11 ms and an average waiting time of 5.5 ms \cite{Liu2018}, while a screen with a refresh rate of 120 Hz has a maximum waiting time of approximately 8.33 ms an average waiting time of 4.165 ms \cite{Huawei2016}. 

Overall, the maximum MTP latency varies based on the types of XR applications and the stages they are in. Full details of each phase/stage mentioned here could be found in Section \ref{sec:capacity_relatedwork}. For instance, in \cite{Mangiante2017}, the maximum MTP latency of 40 ms, 30 ms, 20 ms, and 10 ms are proposed for the early stage, the entry-level VR stage, the advanced VR stage, and the extreme VR stage, respectively. On the other hand, for strong-interaction VR services, the authors of \cite{Hu2020} and \cite{HuaweiTechnologies2016} suggests the maximum MTP latency of 10 ms for the pre-VR and entry-level VR stages, and a more stringent MTP latency of 5 ms for the advanced and ultimate VR stages. For the human perception stage, the authors of \cite{Hu2020} suggest 10 ms as a typical MTP latency for strong-interaction VR services. For weak-interaction VR services and 2D videos, Huawei \cite{HuaweiTechnologies2016} suggests a typical MTP of 30 ms for both pre-VR and entry-level VR stages. On the other hand, weak-interaction VR services and 3D videos need the MTP latency of 20 ms, 20 ms, and 10 ms \cite{HuaweiTechnologies2016} for entry-level VR, advanced VR, and ultimate VR stages, respectively. 

\begin{figure}[htpb!] 
    \centering 
\includegraphics[scale=0.43]{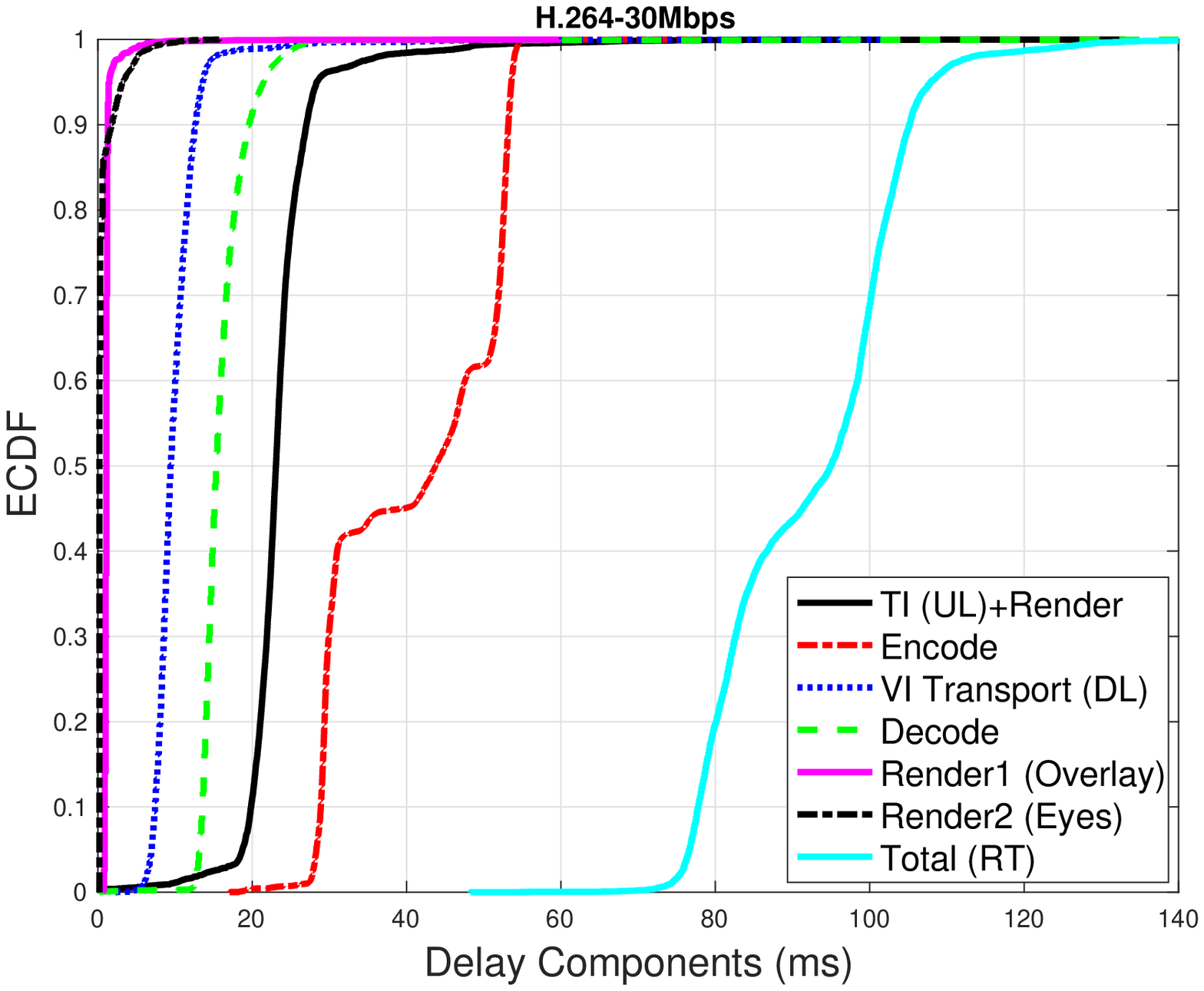} 
\includegraphics[scale=0.43]{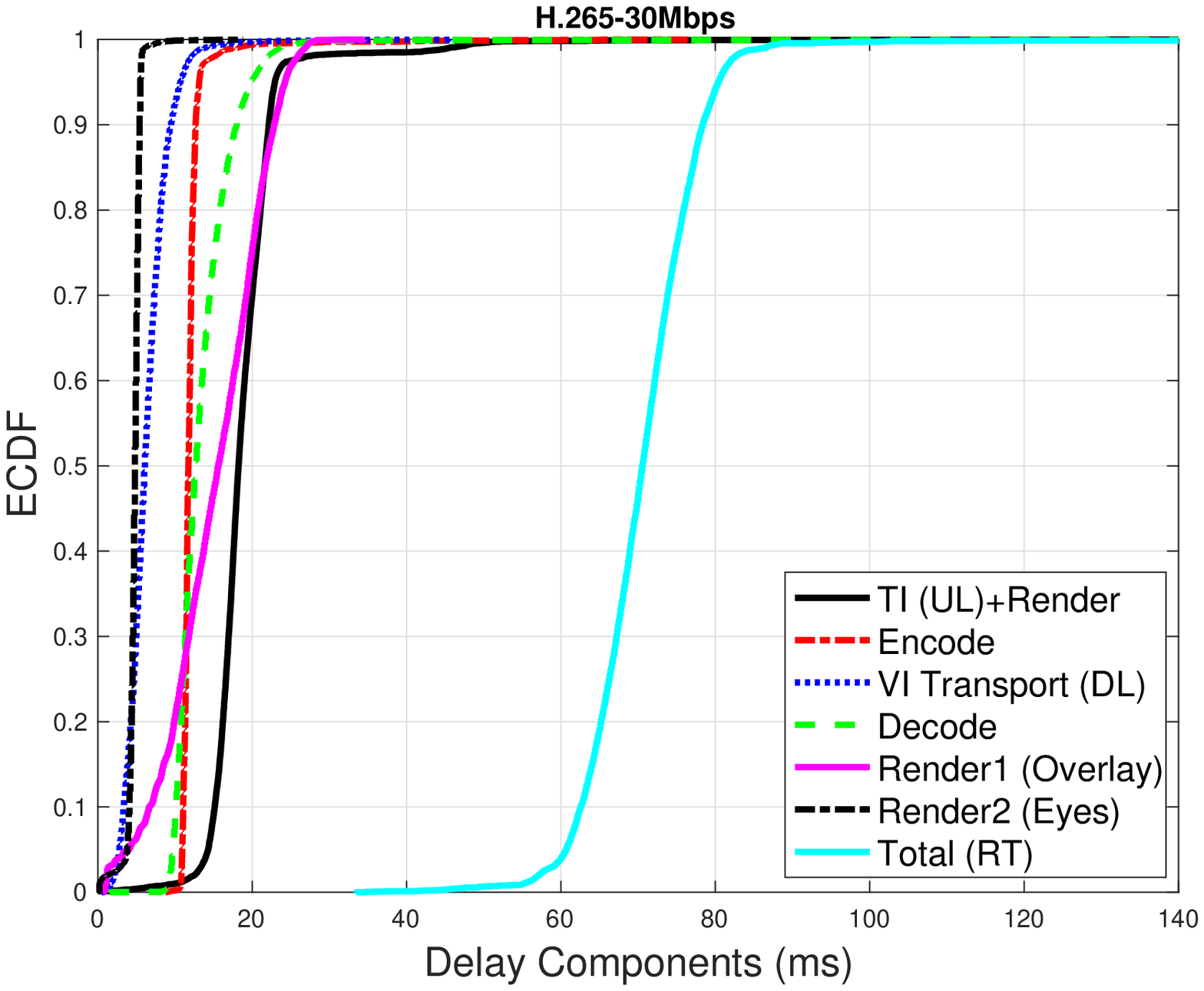}
    \caption{Delay components of different encoders with same target bit rate for 90 seconds of playing Rec Room} 
    \label{fig:delay_H264_RR} 
\end{figure}

Similar to \cite{Hu2020,HuaweiTechnologies2016}, MTP latencies of 20 ms and 10 ms are recommended for weak-interaction and high-interaction VR services, respectively, for the ideal eye-like VR experience. For practical VR experience with Oculus Quest, in \cite{Salehi2020TrafficNetworks}, we had measured the MTP latency for Quest 1 using ALVR while playing Rec Room (RR), a free game that could be played from a server over a wireless link. Two encoding standards (H.264 and H.265) were used with a target bit rate of 30 Mbps to measure the delay components for 90 seconds of playing. The delay components are shown in Fig. \ref{fig:delay_H264_RR}. Quest 1 gathers the tracking information (TI) frequently (50$\sim$60 times per second) to determine the user's position, velocity, and head orientation, and then transmits the information to the server (\underline{{\em UL}}). Based on this information, the server \underline{{\em renders}} the appropriate new video frame. After rendering the new video frame, the server \underline{{\em encodes}} it and sends it back to Quest 1 (\underline{{\em DL}}). Quest 1, then, {\underline{\em decodes}} the received video frame and {\underline{\em renders}} the overlay (the updated haptics positions) on the received video frame. After that, Quest 1 submits the rendered video frame to the display lens(es). The round trip (RT) delay is measured from the time Quest 1 transmits TI to the server until it displays the newly rendered video frame. As seen in Fig. \ref{fig:delay_H264_RR}, the overall delay is lower when H.265 is used for encoding with an average of $\sim$69 ms compared to H.264.

\subsection{\bf Computation in HDM-Edge-Cloud Continuum}

One critical design of any XR experience is to decide where the computation (such as video segments rendering and encoding) takes place. This decision is usually affected by the types of XR applications and the resolution of the corresponding XR videos. One possible option is to perform computation locally inside the HMD, which eliminates any communication delay. However, due to HMD's constrained computing power and battery energy, the supported XR applications are usually less interactive and/or with low resolution compared with other alternatives. Another option is to offload the computation to an external server through a wired connection (USB or HDMI). Although this option increases the available computing power for XR applications without incurring a significant communication delay, wired connection limits the user's mobility, which is a huge disadvantage. 

Yet another option is to offload the computing tasks wirelessly from the HMD. This option acquires the needed computing power without limiting the user's mobility, while keeping the size, the weight, and the energy consumption of HMD small. However, it incurs the additional communication delay in both uplink and downlink, which varies depending on the type of applications.
Unlike AR and MR applications, where they incur high transmission delay in the uplink, VR incurs unnoticeable transmission delay in the uplink but suffers significant transmission delay in the downlink.

Furthermore, with respect to the location of the computing resources, with cloud computing, where the resources reside far away from the user, the roundtrip end-to-end delay measured by ping ranges between 50 ms to 100 ms \cite{ABIResearchQualcomm2017}. When the resources are pushed into telco's cloud, the ping-measured roundtrip delay is reduced to the range between 20 ms to 50 ms \cite{ABIResearchQualcomm2017,Elbamby2018} as shown in Fig. \ref{fig:ARVR}. In edge computing, a.k.a. mobile edge computing (MEC), the ping-measured roundtrip (over-the-air) delay is further reduced to be between 1 ms and 2 ms over 5G \cite{Huawei2016,ABIResearchQualcomm2017,Elbamby2018}. Since these delays are measured by ping, they represent only the communication delay. However, as shown in Fig. \ref{fig:delayComponents} and discussed before, for XR applications, MTP consists of other delay components such as the HMD delay and the computing delay. In particular, the computing delay contributes a large portion of the overall MTP delay and can reach up to 100 ms \cite{Elbamby2018}. Moreover, based on Huawei's test of its Cloud X service over 5G \cite{Zhang2018}, the communication delay contributes only 17.9 ms out of the overall 82.2 ms MTP for VR. Even with the user movement prediction, the communication delay contributes 30\% of the overall MTP \cite{Zhang2018}.

Offloading computing tasks of VR results in a three-way tradeoff among compute, latency, and bandwidth as depicted by a triangle in \cite{Mangiante2017}. In this tradeoff triangle, out of the three angles, at most two can be minimized at the same time. For example,
minimizing the use of computing resources and/or bandwidth incurs more latency. 
Similarly, minimizing the latency requires increasing either the bandwidth or the computing power, or both. 

With respect to the type of XR applications, it is safe to rely on cloud computing for {\em non-interactive} XR applications, such as 360-degree video streaming. These types of XR applications, 
are less susceptible to network latency for the reason that they can cache pre-rendered video segments with a long prediction window to tolerate high communication delay by eliminating on-demand rendering.
Another example that could take advantage of cloud computing (or even the local processing power on HMD) is low-resolution XR applications.
In contrast, interactive XR applications are highly sensitive to delay in general since they are dynamic and less predictable. These applications need to send the interaction information to the computing device, which renders the appropriate video segments based on the information and send them back to the HMD. Such interactivity happens frequently and is difficult to be predicted accurately; therefore, using edge computing, instead of cloud, reduces the overall delay by reducing the communication delay. Another design to satisfy the stringent delay requirements of interactive XR applications is to use hybrid computing. In this design, the computing task is split between two or more computing devices such as local-edge, edge-cloud, and local-edge-cloud. Although hybrid computing has advantages, it introduces complexity to the overall process of XR contents, such as deciding what parts of the video should be processed by which computing device and when. For instance, being able to see a part of the scene or the whole scene in low resolution by using the constrained local computing before the reminder of the scene or the high-resolution version of it comes from the edge (in the local-edge design) is better than seeing nothing (in the edge only design), which is the primary advantage of hybrid computing.

\section{Latency and Bandwidth Minimization for XR}
\label{Sec:Optimizations}

Much work has been conducted to minimize latency and/or bandwidth of XR applications to meet their stringent requirements. The majority of the work targets frame pre-rendering, pre-rendered video segment caching, user movement prediction, multi-resolution video, rate adaptation, and MEC resource allocation. 

The rest of this section categorizes the work based on the minimization approaches used to reduce communication latency, computation latency, or required bandwidth. 

\subsection{\bf Optimization-based approaches}

The authors of \cite{Qian2019} adjust the quality of the streamed volumetric videos based on the quality of the channel and the resulting transmission rate to maintain a specific target delay, which was discussed in Section \ref{Sec:VolumetricVideos}. Other efforts reduce the latency by incorporating such delay in the resource allocation decision of MEC. For instance,
by exploiting the insights that AR applications of different users in proximity share some of the computational tasks, input data, and output data, \cite{Bohez2013,Verbelen2013} form joint optimization problems of optimizing the computation and communication resources to reduce the overall latency. By taking advantage of the same insights,
\cite{Al-Shuwaili2017} proposed Successive Convex Approximation (SCA) to minimize the total energy expenditure of the mobiles {\em for} offloading AR applications' tasks to the edge subject to communication and computation latency constraints. 

\subsection{\bf Parallel transmission and computation based approach}

In this category, the authors of \cite{Liu2018} reduce the overall end-to-end delay ($T_{e2e}$) of Eq. \ref{eq:e2elatency} by minimizing $T_{streams}$ and $T_{display}$. Out of the overall $T_{e2e}$ budget between 20 $ms$ to 25 $ms$, $T_{sense}$ delay takes about 400 $\mu s$ in their WiGig network, and $T_{render}$ takes between 5 $ms$ and 11 $ms$ depending on the tasks' load in their open remote rendering platform. Moreover, they use a display with a 90Hz refresh rate with an average waiting time of 5.5 $ms$ as discussed earlier. The delays mentioned add up to 10.9 $ms$ -- 16.9 $ms$, which leaves less than 10 $ms$ for $T_{stream}$. To reduce the rendering and streaming delay, they propose the Parallel Rendering and Streaming (PRS) approach to parallelize frame rendering and streaming (encoding, transmitting, and decoding). Moreover, to reduce the display latency, they propose the Remote VSync Frame Rendering (RVDR) mechanism to synchronize the arrival time of newly rendered frames (from the edge) with the VSync signal time of the display (in the HMD). Based on the results, their system is capable of supporting the current VR applications with a delay of less than 16 $ms$, and promises to support 4K VR applications with a delay of less than 20 $ms$.

\subsection{\bf Pre-rendering based approaches}

For VR video streaming applications, a VR HMD needs to receive specific video segments with dimensions similar to its viewport from a rendering server based on the user's head orientation (for 360-degree videos) or the user's position and head orientation (for volumetric videos). However, real-time rendering, where the rendering server receives the user's position and/or orientation information and renders the appropriate viewport, consumes the MTP delay budget in VR systems. Therefore, one common technique to reduce such latency and the required throughput is video segments pre-rendering and caching. However, to maximize the potential of the technique, a good design should carefully decide factors such as what parts of the video should be pre-rendered and in which resolution, how long a pre-rendered segment should be cached, and the location of the cache. 

\subsubsection{\bf Retrieval-based approaches}

Work of \cite{Boos2016,Liu2019a} pre-renders all possible scenes, saves them either locally \cite{Boos2016} or in a server \cite{Liu2019a}, and retrieve the desired scene for display. In \cite{Boos2016}, the authors introduce a new VR HMD design termed FLASHBACK, which pre-renders and caches locally all possible scenes a user may view in the HMD to avoid any real-time rendering. Whenever the viewer needs a new scene, the pre-rendered scene will be searched in the local cache and retrieved. To retrieve a specific pre-rendered scene from the cache, a hierarchical index is built. Based on the evaluation, FLASHBACK is appropriate for both static and dynamic scenes, where it increases the number of frames per second by eight folds and decreases the latency and power consumption by 15 and 97 folds, respectively, compared with a mobile local-rendering scenario. 

On the other hand, the authors of \cite{Liu2019a} designs a cross-layer approach termed Chord that considers optimizations at both the application layer and wireless network. In the application layer, Chord performs offline pre-rendering to save the computing time, differentiation between interactive and non-interactive objects in the scenes to render each type in different quality, Adaptive Content Quality Control (AQC) based on the information from the lower layers to consider the different network conditions of each user, and prediction of user movements to send the scenes that are most likely to be seen by the user. In the wireless network, Chord estimates the accurate throughput based on the information from lower layers to use it in AQC, minimizes the interference by carefully picking MU-MIMO based on channel correlation, and reduces the uplink delay by using three AP-based solutions. It is worth mentioning that this work, in addition to \cite{Qian2019} which was discussed in Section \ref{Sec:VolumetricVideos}, is one of the few efforts that dealt with volumetric videos (6DoF) instead of 360-degree videos (3DoF).

\subsubsection{\bf Viewport-adaptive techniques}
\label{Sec:Viewport-adaptive}
\begin{figure*}[htpb!] 
    \centering 
\includegraphics[width=0.9\linewidth]{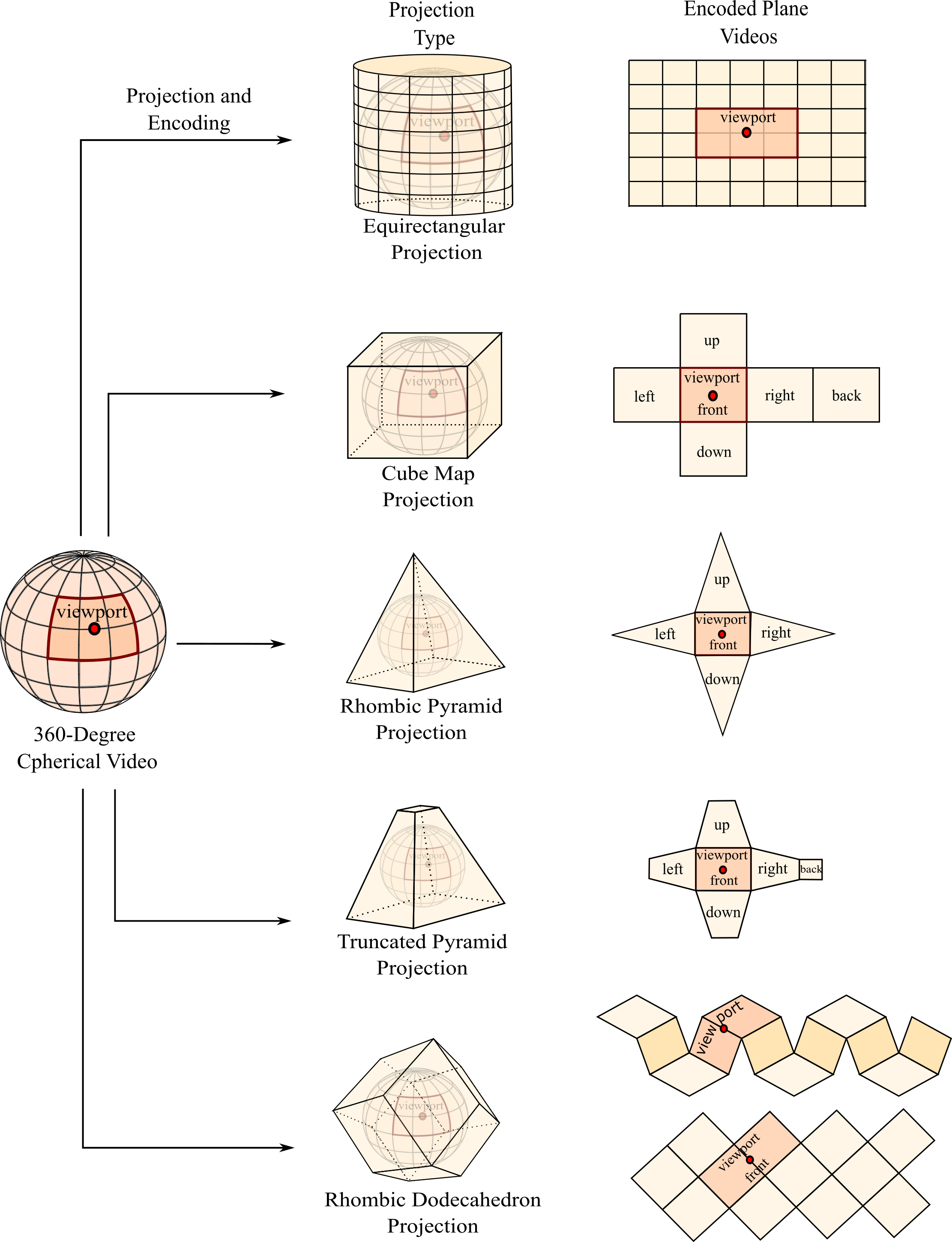} 
    \caption{360-degree video projections} 
    \label{fig:projections} 
\end{figure*}

Another approach to reducing latency is to use the adaptive bitrate streaming (ABS) technique which helps handling the bitrate fluctuation (between users and for a user over time). Regular video streaming has used this approach for a long time to adapt the resolution of videos based on the quality of the users' connection. However, applying such technique directly to 360-degree videos is not feasible because of their spherical nature. Therefore, projecting spherical videos to plane videos becomes necessary before streaming since the existing video encoders cannot handle spherical videos. 

There are many projection approaches to map 360-degree frames into plane videos \cite{Corbillon2017, Kammachi-Sreedhar2017}, including equirectangular projection, cube map projection, rhombic pyramid projection, truncated pyramid projection, and rhombic dodecahedron projection. Fig. \ref{fig:projections} depicts how a single frame of 360-degree video is converted to a frame of plane video by using each of these project schemes. Some of these projections, e.g., equirectangular, suffer from oversampling, where a single pixel in the sphere is mapped to a pair of pixels in the plane projection \cite{Corbillon2017}. The amount of duplicated pixels in equirectangular is estimated to be 30\% \cite{Chakareski2018}. On the other hand, others, such as the rhombic pyramid projection, suffer from undersampling, where some pairs of pixels in the sphere are projected to a single pixel in the plane, resulting in information loss and distortion \cite{Corbillon2017}. In general, the performance of the multi-resolution cube map \cite{Corbillon2017, Kammachi-Sreedhar2017} and equirectangular projections outperform other \cite{Kammachi-Sreedhar2017}.

\begin{figure*}[htpb!] 
    \centering 
\includegraphics[width=0.9\linewidth]{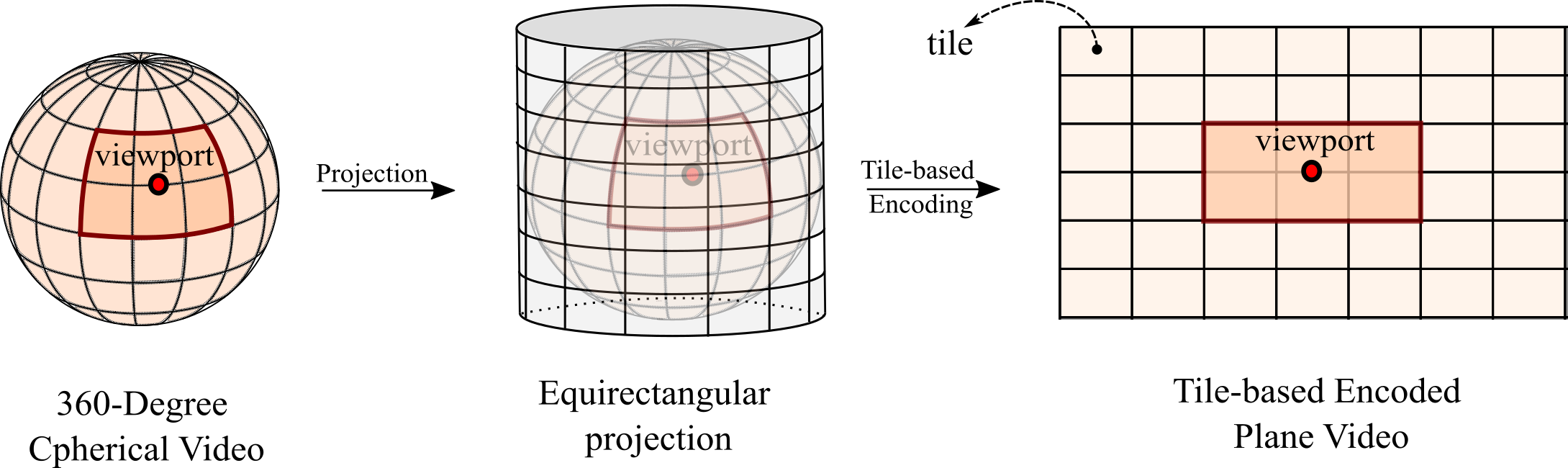} 
    \caption{Converting a frame of 360-degree video into a frame of tile-based plane video} 
    \label{fig:tileBased} 
\end{figure*}

After mapping 360-degree videos into plane videos, the whole plane video could be encoded in different resolutions and streamed based on the user's available bit rate. However, the users can see approximately 20\% of the mapped plane video \cite{Bao2016ShootingVideos}, which is equivalent to the viewport of their HMD. This viewport is determined by the HMD's FoV and the user's head orientation. Moreover, the mapped plane videos consume more bandwidth than regular videos with a similar resolution. In other words, if the resolution of the viewport of the mapped plane 360-degree video equals the resolution of a regular video, the whole mapped plane video will consume 4-6 times more bandwidth than regular videos \cite{Bao2016ShootingVideos}. Therefore, viewport-adaptive techniques are desired, where the streaming process is determined by the expected viewport. These techniques could be categorized into {\em projection}-based viewport-adaptive techniques \cite{Corbillon2017,Hu2019,Kammachi-Sreedhar2017} and {\em tile}-based viewport-adaptive techniques \cite{Aksu2018,Bao2016ShootingVideos,Chakareski2018, Fan2017FixationReality,Fu2019,Ghosh2017,Ghosh2018, Guntur2012OnLAN,Jiang2020,Kan2019DeepStreaming, Kimata2012MobileSystem,Krouka2020, Mavlankar2009Pre-fetchingSequences, Mavlankar2010VideoPan/Tilt/Zoom, Petrangeli2017AnVideos,Qian2016,Qian2018Flare:Devices, Sassatelli2019,Zhang2019DRL360:Learning}. Transmitting only the viewport part of the video helps to reduce the consumed bandwidth by approximately 80\%. 

\noindent
\phantom{2x} {\bf({\em a}) Projection-based}

For the projection-based viewport-adaptive technique, the viewport is determined by the faces resulting from the projections, such as the lower four projections in Fig. \ref{fig:projections}. Then, the rendering server transmits the face of the projection that corresponds to the viewport in the highest possible resolution.

\noindent
\phantom{2x} {\bf({\em b}) Tile-based}

In contrast, for the tile-based viewport-adaptive technique, the rendering server splits the 360-degree video into chunks, each of which consists of multiple spherical frames. Each frame is projected using equirectangular projection into a plane frame which is later divided into multiple spatial areas called tiles as shown in Fig. \ref{fig:tileBased}. The tiles representing the viewport are transmitted in the highest possible resolution based on the quality of the users' connection.

Comparing with applying the ABS technique to regular videos, the tile-based viewport-adaptive technique is not trivial. Each plane frame has to have representations in multiple resolution  corresponding to the bitrate levels of the users' connections. For each representation, the collection of tiles that corresponds to the current viewport has to be selected. With $L$ levels of bitrates and $T$ tiles in each frame, the solution space will have $L^T$ maximum choices for each frame, which leads to a combinatorial explosion. Moreover, to maximize the potential benefit of viewport-based ABS techniques, accurate viewport prediction and bandwidth estimation methods are essential. 

For viewport prediction, users' head movements could be predicted with an accuracy of more than 90\% even with simple methods such as linear regression \cite{Qian2016}. However, there is a tradeoff between prediction accuracy and prediction window length. For instance, in \cite{Bao2016ShootingVideos}, the authors were able to accurately predict head movements (and hence viewport prediction) up to a 500-$ms$ window. In contrast, for a 2.5-second prediction window, the accuracy of all approaches in \cite{Nguyen2018YourPrediction} degrades to less than 50\%. Therefore, another critical component of a successful viewport-based ABS scheme is to have a  buffer with a good replacement strategy, such as \cite{Hu2019}, to keep the predicted viewports (that had been pre-rendered) up to date. 

Specifically, viewport prediction is achieved by predicting the user head movements based on information such as the current and past head rotations \cite{Bao2016ShootingVideos, Chakareski2018, Qian2018Flare:Devices}, content-related information \cite{Fan2017FixationReality, Mavlankar2009Pre-fetchingSequences, Mavlankar2010VideoPan/Tilt/Zoom, Nguyen2018YourPrediction}, and/or crowdsourced information (from multiple users) \cite{Ghosh2018, Guntur2012OnLAN, Liu2019}. The prediction mechanisms that utilize the information to predict upcoming viewports include heuristic algorithms and/or probabilistic approaches \cite{Chakareski2018,Ghosh2017,Ghosh2018,Hu2019}. In addition, machine learning (ML) based solutions become effective, such as linear regression (LR) \cite{Bao2016ShootingVideos, Qian2018Flare:Devices}, ridge regression (RR) \cite{Qian2018Flare:Devices}, support vector regression (SVR) \cite{Qian2018Flare:Devices}, neural network (NN) \cite{Aksu2018, Bao2016ShootingVideos}, recurrent neural network (RNN) (specifically, Long Short Term Memory [LSTM]) \cite{Fan2017FixationReality}, reinforcement learning (RL) \cite{Jiang2020}, deep reinforcement learning (DRL) \cite{Kan2019DeepStreaming, Krouka2020, Sassatelli2019, Zhang2019DRL360:Learning}, and sequential deep reinforcement learning (SDRL) \cite{Fu2019}. According to the evaluations in \cite{Mao2017NeuralPensieve}, DRL-based approaches outperform other approaches. Moreover, it was shown in  \cite{Fan2017FixationReality, Nguyen2018YourPrediction, Sassatelli2019, Xu2018GazeVideos, Zhang2019DRL360:Learning} that deep learning, specifically LSTM, yields the best results for viewport prediction in VR applications. 

\subsubsection{\bf Considerations and technologies for transmitting and displaying pre-rendered contents}

After predicting the viewport, the rendering server has to decide what to transmit and in which resolution. Some work decided to submit only the tiles that compose the expected viewports out of the entire plane frame \cite{Bao2016ShootingVideos, Fan2017FixationReality, Hu2019, Liu2019, Qian2016, Qian2018Flare:Devices}. With perfect prediction accuracy, it will save approximately 80\% in consumed bandwidth, as mentioned earlier. However, when the the prediction is not accurate, the user will see a blank screen for a moment, which affects the perceived QoE and, worse, it will cause motion sickness when it occurs frequently. Others send the viewport in the highest possible resolution, based on the user's available bandwidth, and the rest of the region in low resolution \cite{Aksu2018,Chakareski2018,Corbillon2017,Fu2019,Ghosh2017,Ghosh2018,Jiang2020,Kammachi-Sreedhar2017,Kan2019DeepStreaming,Krouka2020,Petrangeli2017AnVideos,Sassatelli2019,Zhang2019DRL360:Learning}. This approach mitigates motion sickness caused by inaccurate prediction, since the viewer is going to see the real viewport of the video in low resolution, instead of a blank screen, until its high resolution version comes. In addition, it will conserve bandwidth, compared with sending the entire 350-degree frame in the highest possible resolution. 

However, frequently jumping from low resolution to high resolution also affects the perceived QoE. Moreover, sending the remainder of the video frame even in low resolution, while some parts of it are improbable to be seen (such as the area in the opposite direction of the user's viewport) is considered a waste of bandwidth. Therefore, a better solution is to submit the expected viewport in the highest possible resolution and the remainder of the video frame in a resolution corresponding to the likelihood of being seen. In other words, if the area behind the user is unlikely to be seen, it will be sent in a very low resolution or not be sent at all.

In terms of the technologies used for displaying pre-rendered contents, some work uses mobile devices to display the 360-degree video \cite{Liu2019,Qian2018Flare:Devices} and the orientation of the mobile device determines the viewport of the video. Regarding the technologies used for transmitting pre-rendered contents, HTTP and HTTP/2 are used in the Dynamic Adaptive Streaming over HTTP (DASH) protocol \cite{Liu2019,Petrangeli2017AnVideos,Qian2018Flare:Devices} to transmit the encoded video segments from the server to the client. Then, the client uses their browsers \cite{Liu2019}, mobile devices \cite{Qian2018Flare:Devices}, or HMDs to display the encoded video segments. 

Moreover, with respect to the type of videos, most work targets on-demand 360-degree video streaming, where the 360-degree video is located in a server, such as YouTube and Facebook, and the users access it whenever they want. Another type is {\em live} 360-degree video streaming \cite{Aksu2018,Kimata2012MobileSystem,Liu2019,Mavlankar2009Pre-fetchingSequences}, where an event (such as a concert) is recorded in real-time as a 360-degree video using an omni-directional camera and then streamed to multiple users at the same time. These live videos are more challenging than on-demand videos because there is no pre-rendering in advance due to the lack of users' head movement history. In other words, simultaneous interactions with all the on-line users need to be handled to encode the video upon receiving their respective heads movement feedback in real-time. The matter is further complicated by the trade-off between the latency and the perceived video quality.

For the volumetric videos \cite{Liu2019a,Qian2019}, where there are six degrees of freedom (6 DoF) instead of three in the 360-degree videos, they are generally more challenging than 360-degree videos since both head rotational movements and body translational movements need to be predicted. In \cite{Liu2019}, the authors split the prediction task of each movement and handle it as a separate prediction task. Detailed information about volumetric videos was discussed in Section \ref{Sec:VolumetricVideos}. 

Table \ref{Table:currentResearch} categorized and summarized the reviewed research described in this section.
\section{The Reliability Requirement of XR}


\label{Sec:Reliability}

In addition to high capacity and low latency, immersive XR experiences require their connections to be ultra-reliable, which means XR contents are to be delivered on time and with ultra-low packet loss. Immersive XR experiences also demand the connections to be stable and not fluctuant. In other words, the connection has to keep the same level of performance (in terms of capacity and latency) during the entire session with an ultra-high frame delivery success rate. In \cite{ABIResearchQualcomm2017}, this requirement is referred as \textit{uniform experience}. 
However, maintaining uniform connection performance for all the users during their entire sessions is challenging, especially for high mobility use cases. For instance, varying connection bandwidth, addressed via dynamic rate adaption, leads to varying resolution \cite{ABIResearchQualcomm2017}, which, coupled with latency variation, results in fluctuating XR experience or even motion sickness.

There are two approaches to achieving an ultra-reliable connection: simultaneous multi-connectivity (MC) \cite{ABIResearchQualcomm2017,Elbamby2018} and massive multiple input multiple output (MIMO) \cite{ABIResearchQualcomm2017}. MC's main objective is to improve the connection's data rates and reliability, i.e., achieve uniform experience by using a diversity of connections. These connections could be established using the same carrier frequencies (intra-frequency MC) or different carrier frequencies (inter-frequency MC) \cite{Elbamby2018}. An example of the intra-frequency MC is single-frequency networks (SFNs), where several transmitters use the same frequency to send the same signal to the same user jointly. In inter-frequency MC, a transmitter (or several transmitters) uses multiple frequencies simultaneously to send the same signal. An example of inter-frequency MC is when a transmitter is connected to multiple 5G/4G base stations and Wi-Fi access points simultaneously to increase the connection reliability. Although MC enhances reliability significantly, it comes with the major drawbacks of more resource consumption and additional delays \cite{Elbamby2018}. When a transmitter utilizes multiple resources, it will negatively impact the delay experienced by the remaining transmitters. Moreover, the use of re-transmission and redundancy at the PHY layer will cause additional delay \cite{Elbamby2018}. On the other hand, in massive MIMO, multiple antennas are used to direct the signal to the user, which increases the connection's uniformity and capacity \cite{ABIResearchQualcomm2017}.

Many papers have discussed the minimum required frame delivery success rate or the maximum packet loss rate to achieve a satisfying immersive XR experiment. The maximum packet error rate (PER) for immersive VR experiments, according to \cite{Elbamby2018}, is similar to what is defined by 3GPP, which is $10^{-5}$. This number can be translated to a minimum of five-nines (99.999\%) packet delivery success rate. 
Out of the four Huawei stages \cite{HuaweiTechnologies2016}, which are described in Section \ref{sec:capacity_relatedwork}, they propose packet loss rates of 2.40$\times 10^{-4}$ and 2.40$\times 10^{-5}$ for the pre-VR and entry-level VR stages in the case of weak-interaction VR services, which translate to 99.976\% and 99.9976\% packet delivery success rates,  respectively. Furthermore, they specify more stringent reliability requirements with a typical packet loss rate of $10^{-6}$, which translates to a minimum of six-nines (99.9999\%) packet delivery success rate, for the advanced VR and ultimate VR in the case of weak-interaction VR services and for all of their aforementioned VR stages in the case of strong-interaction VR services. Similarly, the authors of \cite{Hu2020} specify a similar packet loss rate of $10^{-6}$ for all of their VR phases mentioned in Section \ref{sec:capacity_relatedwork}. Generally speaking, \cite{Adame2020} describes any connection with a packet delivery success rate greater than three-nines (99.9\%) and four-nines (99.99\%) to be adequate for the existing VR and AR applications, respectively. These numbers translate to a maximum packet loss rates of $10^{-3}$ and $10^{-4}$ for the existing VR and AR applications, respectively. For nowadays AR applications, the connection jitter has to be in the range of a few microseconds \cite{Nasrallah2019}.

For the current VR stage in the case of weak-interaction VR services, TCP protocol is adequate to transmit VR traffic. Therefore, to compute the maximum allowable packet loss rate ($L$), we use the following formula.
\begin{equation}
\label{eq:packetloss}
L \leq (\frac{MSS}{throughput \times RTT})^2 ,
\end{equation}
where $MSS$\footnote{$MSS$ is usually equal to 1460 bytes (11680 bits).} is the maximum segment size in bits, {\em throughput} is the required throughput in bps, and $RTT$ is the round-trip time (MTP delay) in seconds. For instance, the comfortable-experience VR phase in \cite{HuaweiiLab2017} has a recommended MTP of 20 ms and a required throughput of 140 Mbps to transmit the full-view video. Therefore, the packet loss rate has to be $\leq (\frac{1460\times8}{140\times10^6\times20\times10^{-3}})^2\approx 1.7\times10^{-5}$, which translates to packet delivery success rate of $(1-1.7\times10^{-5})\times100 \approx99.9983\%$.

For strong-interaction VR services, UDP protocol is recommended for transmitting VR traffic \cite{HuaweiiLab2017}. Since UDP does not support re-transmission of lost packets, which make it more susceptible to packet loss than TCP, a requirement of lower packet loss rate is recommended for strong-interaction VR services. Typically, a packet loss rate of less than $10^{-6}$ is adequate for these VR applications \cite{Hu2020,HuaweiTechnologies2016,HuaweiiLab2017}. 

Therefore, similarly, the ideal eye-like VR, one of the experiences considered in this paper, should have a packet loss rate of $10^{-6}$ (more than 99.9999\% packet delivery success rate). On the other hand, the approximate maximum packet loss rate for Quest 1 could be computed using Eq. \ref{eq:packetloss} and the information provided in Table \ref{Table:Quest2Information} and Fig. \ref{fig:delay_H264_RR}. For instance, the required throughput to transmit the full-view video is 62.85 Mbps when the refresh rate is set to 72Hz and the average total RT delay is 69 ms when the H.265 encoding standard is used. From Eq. \ref{eq:packetloss}, the recommended packet loss rate should be less than or equal $7.2\times 10^{-6}$ (more than 99.99928\% packet delivery success rate). Table \ref{Table:SummaryofTheTwoExperiences} summarizes the requirements of the two VR experiences considered in this paper.

{\renewcommand{\arraystretch}{1.43}
\begin{table*}[ht]
\centering
\begin{tabular}{||c||c|c||c|c||}\hline\hline
\multicolumn{3}{||c||}{\backslashbox{\textbf{Requirement}}{\textbf{VR Experience}}} & \thead{\textbf{Quest VR Experience}} & \thead{\textbf{Ideal Eye-like VR Experince}}\\
      \hline\hline
\multicolumn{3}{||c||}{\textbf{Full-view Video Resolution}} & 6770$\times$3380 & 72,000$\times$36,000 \\\hline
\multicolumn{3}{||c||}{\textbf{Single Eye Resolution\tablefootnote{This value is computed as single eye {\em FoV}$\times ppd$}}} & 1824$\times$1840 & 31,000$\times$26,000 \\\hline
\multicolumn{3}{||c||}{\textbf{Single Eye {\em FoV$_{(h)}$$\times$ FoV$_{(v)}$}}} & 97$\degree$$\times$98$\degree$ & 155$\degree$$\times$130$\degree$\\\hline
\multicolumn{3}{||c||}{\textbf{Bit per Color ({\em bpc})}} & 8 & 8\\\hline 
\multicolumn{3}{||c||}{\textbf{Bit per Pixel ({\em b$_{depth}$\tablefootnote{This value is computed under the assumption that no chroma subsampling is used})}}} & 24 & 24 \\\hline
\multicolumn{3}{||c||}{\textbf{Pixel per Degree ({\em ppd})}} & 18.8 & 200 \\\hline
\multicolumn{3}{||c||}{\textbf{Refresh Rate}} & 72 Hz & 77 Hz \\\hline\hline
\multirow{6}{*}{\begin{turn}{90}\textbf{Service Requirements}\end{turn}} & \multirow{3}{*}{\begin{turn}{90}\textbf{Bit Rate\tablefootnote{This is the bit rate required to transmit the {\em FoV}}}\end{turn}}& \textbf{No compression} & $\leqslant$ 10.80 Gbps & $\leqslant$ 2.71 Tbps \\ \cline{3-5}
 & & \textbf{Low Latency Compression (20:1)} & $\leqslant$ 553.08 Mbps & $\leqslant$ 138.72 Gbps \\ \cline{3-5}
 & & \textbf{Lossy Compression (600:1)} & $\leqslant$ 18.44 Mbps & $\leqslant$ 4.62 Gbps \\ \cline{2-5}
 & \multicolumn{2}{c||}{\textbf{Typical MTP Latency\tablefootnote{For Quest experience, this latency is computed when H.265 encoding standard is used}}} & $\leqslant$ 69 $ms$ & $\leqslant$ 20 $ms$ \\ \cline{2-5}
 & \multicolumn{2}{c||}{\textbf{Typical Packet Loss Rate}} & $\leqslant 7.2\times10^{-6}$ & $\leqslant 1\times10^{-6}$ \\ \cline{2-5}
 & \multicolumn{2}{c||}{\textbf{Typical Packet Delivery Success Rate}} & $\geqslant$ 99.99928\% & $\geqslant$ 99.9999\% \\\hline\hline
\end{tabular}
\captionof{table}{Summary of the QoS requirements for the two VR experiences}
\label{Table:SummaryofTheTwoExperiences}
\end{table*}
}


\section{Conclusion}
\label{Sec:Conclusion}
This paper discussed the three main requirements for any satisfying immersive XR experience: high capacity, ultra-low latency and reliability, and the ultra-low latency is further composed of low computing latency and low communication latency. For each requirement, there are technology enablers. For instance, mmWave and the coming IEEE 802.11be Extremely High Throughput (EHT) or Wi-Fi 7 increase capacity. Also, edge computing and proactive computing and caching reduce latency. Moreover, multi-connectivity (MC) and massive MIMO achieve ultra-high reliability, and the technique of multicast improves scalability and may reduce communication latency.

Considerable studies have been done to reduce the required delay and/or bit rate of XR applications by optimizing one or more aspects of C3 \cite{Elbamby2018} (caching, computing, and communication). The main objective (direction) of these studies is to decide: (1) how to transmit (e.g., technology used, multi- vs. unicast, and channel access mechanism) and when to transmit (scheduling), (2) what to transmit and in which quality (prediction, multilayers), and (3) how to compute (real-time rendering vs. pre-rendering) and where to compute and/or store (locally, in the edge, in the cloud, or hybrid) the XR traffic. Even though these studies helped meeting the current XR requirements, more work needs to be done in the C3 aspects of XR applications to satisfying future XR requirements, such as human eye-like XR experience. Moreover, out of the three XR requirements, focusing on improving only one factor, such as the latency (computing and/or communication latency), would be wise since improving one factor may relax the other two requirements. In other words, if an approach manages to halve the latency, the lax required bit rate could be significantly reduced. Moreover, minimizing the latency offers an additional window for re-transmission when a packet loss occurs, which relaxes the reliability requirement (in terms of packet loss). 

By taking the emerging Wi-Fi standard 802.11be (Wi-Fi 7) as an example, with its expected high throughput of 30 Gpbs and features like multi-link operation and multi-AP coordination, XR traffic could be better supported. However, to optimize the performance, joint transmission scheduling and multi-AP coordination need to be further investigated. In addition, the multi-link operation of 802.11be (including multi-band aggregation, multi-band and multi-channel full duplex, and data and control plane separation) could be better utilized to increase the throughput and/or the network's reliability for XR applications.


\bibliographystyle{plain} 
\bibliography{references.bib} 

\end{document}